\documentclass[fleqn,usenatbib]{mnras}

\usepackage{newtxtext,newtxmath}

\usepackage[T1]{fontenc}

\DeclareRobustCommand{\VAN}[3]{#2}
\let\VANthebibliography\thebibliography
\def\thebibliography{\DeclareRobustCommand{\VAN}[3]{##3}\VANthebibliography}

\usepackage{svg}
\usepackage{graphicx}	
\usepackage{amsmath}	
\usepackage{subcaption}
\usepackage{courier}

\newcommand{\Msun}{\mbox{\,$M_{\odot}$}}        

\title[Evolution of very massive stars]{Mass-loss implementation and temperature evolution of very massive stars}

\author[G N Sabhahit et al.]{
Gautham N. Sabhahit$^{1}$\thanks{E-mail: gautham.sabhahit@armagh.ac.uk},
Jorick S. Vink$^{1}$,
Erin R. Higgins$^{1}$\&
Andreas A.C. Sander$^{1,2}$\\
$^{1}$Armagh Observatory and Planetarium, College Hill, Armagh BT61 9DG, N. Ireland\\
$^{2}$Zentrum f{\"u}r Astronomie der Universit{\"a}t Heidelberg,
Astronomisches Rechen-Institut, M{\"o}nchhofstr. 12-14, 69120
Heidelberg, Germany}

\date{Accepted XXX. Received YYY}
\pubyear{2022}

\begin{document}
\label{firstpage}
\pagerange{\pageref{firstpage}--\pageref{lastpage}}
\maketitle

\begin{abstract}
\noindent
Very massive stars (VMS) dominate the physics of young clusters due to their ionising radiation and extreme stellar winds. It is these winds that determine their lifepaths until expiration.
Observations in the Arches cluster show that VMS all have
similar temperatures. The VLT-Flames Tarantula survey analysed VMS in the 30 Dor region of the LMC also finding a narrow range of temperatures, albeit at higher values -- likely a metallicity effect. 
Using MESA, we study the main-sequence evolution of VMS with a new mass-loss recipe that switches from optically-thin O-star winds to optically-thick Wolf-Rayet type winds through the model-independent transition mass-loss rate of Vink \& Gr\"afener. 
We examine the temperature evolution of VMS with mass loss that scales
with the luminosity-over-mass $(L/M)$ ratio and the Eddington parameter
($\Gamma_\mathrm{e}$), assessing the relevance of the surface hydrogen (H)
abundance which sets the number of free electrons.
We present grids of VMS models at Galactic and LMC metallicity and compare our temperature predictions with empirical results.
Models with a steep $\Gamma_\mathrm{e}-$dependence evolve horizontally in the Hertzsprung-Russel (HR) diagram at nearly constant luminosities, requiring a delicate and unlikely balance between envelope inflation and enhanced mass loss over the entire VMS mass range. 
By contrast, models with a steep $L/M-$dependent mass loss are shown to evolve vertically in the HR-diagram at nearly constant $T_{\rm eff}$, naturally reproducing the narrow range of observed temperatures, as well as the correct trend with metallicity. 
This distinct behavior of a steeply dropping luminosity is a self-regulatory mechanism that keeps temperatures constant during evolution in the HR-diagram.  \\

\end{abstract}




\section{Introduction}

Over the last decade the interest in 
Very Massive Stars (VMS) up to 300\,\Msun\ has grown substantially \citep{Crowther2010, Best2020, Roy2020}. VMS are defined as stellar objects with masses over 100\,\Msun\ \citep{Vink2015}. Their spectral appearance no longer resembles that of an absorption-line dominated O-star spectrum but that of an emission-line dominated Wolf-Rayet (WR) star of the nitrogen (N) type with left-over hydrogen (H) in their spectra. Such WNh stars are likely still on the H-burning main sequence (MS) \citep[e.g.][]{Massey1998, Dekoter1998, Hamann2006, Martins2022}, with enhanced mass-loss rates in comparison to canonical O-type stars \citep{Vink2011, Graf2011,Best2014}.

The evolution of VMS is highly uncertain due to the unknown physics in close proximity to the Eddington limit of radiative pressure. If mass-loss rates are low, such as at low metallicity ($Z$), VMS may produce pair-instability supernovae (PISNe) where the entire star is obliterated \citep{Barkat1967, Fryer2001,UN2002,Scan2005,Langer2007,Kasen2011, Hirschi2015, Woosley2015}, and one such PISN could potentially produce more metals than an entire initial mass function (IMF) below it \citep{Langer2012}. If mass-loss rates are high, such as in the high-$Z$ environment of the Milky Way or Large Magellanic Cloud (LMC), discussed in this paper, VMS may evaporate themselves, largely already on the MS \citep{Belkus2007, Yung2008, Vink2018}.
The black hole (BH) mass function and its maximum mass will depend strongly on the exact formulation of the mass-loss input \citep{Vink2021}.

In addition to increased mass-loss rates, objects close to the Eddington limit may be subjected to envelope inflation \citep{Ishi1999, Petrovic2006, Graf2012, Sanyal2015, Jiang2015}. When stellar evolution modellers employ extrapolated standard O-star mass-loss rates \citep[e.g.][]{Vink2000} into the VMS regime, the rates are so low that the stars inflate to such large radii and low effective temperatures that they are shown to cross the Humphreys-Davidson (HD) limit \citep{HD1979}. In other words the models are contradicted by the observations. For these reasons, stellar modellers generally adapt a criterion to prevent this. In MESA that can be achieved with the MLT++ option \citep{MESA13, Sabhahit2021}. Also the Geneva VMS models by e.g. \citet{Yusof2013} use an additional energy transfer criterion in these radiation-dominated envelopes. 
3D radiation-hydrodynamics computations \citep[e.g.][]{Jiang2015, Schultz2020} do not necessarily agree with MLT++ and related interventions. One plausible alternative to this inflation problem is that O-star extrapolated VMS mass-loss rates are not sufficiently high. When mass-loss rates near the Eddington limit are sufficiently enhanced, the MLT++ intervention is no longer required.

In addition to the HD limit, and the absence of very luminous cool supergiants accurate stellar parameters of VMS have been obtained over the last decade.
Two interesting features are observed from the temperatures of VMS in known clusters that hosts such objects. First, the observed temperatures of VMS are confined to a small range of values: log$(T_\mathrm{eff})\approx 4.5-4.6$ in the Arches cluster \citep{Martins2008} near the Galactic center whose Z content is hinted to be super-solar, log$(T_\mathrm{eff})\approx 4.62$ in the Galactic cluster NGC3603 \citep{Crowther2010}, log$(T_\mathrm{eff})\approx 4.6-4.7$ in the 30 Dor cluster of the LMC \citep{Best2014} with its approximately half-solar $Z$, and log$(T_\mathrm{eff})\approx 4.72$ in the central region of 30 Dor that hosts the brightest VMS in the LMC, the R136 cluster \citep{Crowther2010}. All these objects in their respective clusters have a nearly vertical alignment in the HRD (see Fig. \ref{fig:HRD_gal_TEST} and \ref{fig:HRD_lmc_TEST}). Second, the temperature distribution might suggest a consistent temperature-Z correlation with VMS having higher temperatures at lower $Z$.  The temperature distribution of VMS could indeed be an interplay between the effects of $Z$ and the age of the cluster, but a consistent trend with $Z$ might hint towards a dominant $Z$ effect.

At face value the vertical alignment of the observed temperatures of stars could just be due to a burst of star formation, with the stars in the cluster all at the same age. For typical ages of young stellar clusters between $0-2$ Myr, canonical O stars lie slanted and parallel to the zero age main sequence (ZAMS), which which becomes increasingly vertical for O stars in the mass range 60-100 $M_\odot$. While a constant $T_\mathrm{eff}$ might indeed be interpreted as evidence for a burst of star formation for typical ages of young clusters, the situation is more complex for masses above 100 $M_\odot$ as these VMS are subjected to substantial envelope inflation. This implies that the most massive stars will inflate to lower $T_\mathrm{eff}$ than the less massive stars in the population. Conversely, higher mass loss might reduce and/or offset this effect, but again effects for the most massive star might flip the stars over to much higher $T_\mathrm{eff}$ than for the lower mass stars of the population. This implies that for normal evolution it is quite a challenge to exactly offset the inflation and mass loss effects by exactly the correct balance - over the entire mass and metallicity range. A particular balance between the two effects that might work at a certain initial mass and initial $Z$ fails to do so across other masses and metallicities.



Existing VMS models in the literature that use canonical O-star mass loss rates are unlikely to evolve at similar temperatures for varying initial masses and fail to explain the  observed features mentioned previously. In this paper we explore two different scaling for mass loss assuming different importance for surface H, and how they affect the temperature evolution of VMS. The aim is to have models evolve at nearly constant temperatures without being highly sensitive to the input physics. As we see in Sec. \ref{sec:results}, a drop in luminosity thus suppressing the effects of both inflation and mass loss, has a self-regulatory effect which naturally keeps the temperatures constant throughout the evolution.

The paper is organized as follows. We begin by giving a general overview of the input physics used to model the H burning phase of VMS in Sec. \ref{sec:methods}. The mass loss properties of O and WNh stars, and the transition mass loss that characterizes the switch from optically thin to optically thick wind that we use to calibrate our absolute rates is presented in Sec. \ref{sec:mass_loss}. In the results section (Sec. \ref{sec:results}), we present our grid of VMS models at Galactic and LMC metallicity using our $L/M$-dependent mass loss and explore the chemically homogeneous evolution of VMS for $M_\mathrm{init} \gtrsim 200 M_\odot$. In Sec. \ref{sec:discussion}, we discuss the reasons for the temperature insensitivity of our models and the possible consequences of having a fully mixed star. Our results are summarized in Sec. \ref{sec:conclusion}

\section{Methodology}
\label{sec:methods}

In this section we give a general overview of the different inputs to model the MS evolution of VMS including their initial masses, initial metallicities, overshooting  and rotational efficiencies. The massive star models presented in this paper are produced using the 1D stellar evolution code MESA (version r12115) \citep{MESA11, MESA13, MESA15, MESA17, MESA19}. Unless specified, the models are evolved until the end of core H burning, where the models stop once the central hydrogen mass fraction ($X_\mathrm{c}$) falls below 0.01. We describe the macro- and micro-physics implemented in our models below.

In this paper, we study the MS evolution of massive stars with initial masses ranging from 60 $M_\odot$ to 500 $M_\odot$. We consider two initial metallicities corresponding to the Galaxy and the Large Magellanic Cloud, where we use the physical transition mass loss properties to calibrate the absolute VMS mass-loss rates in our models (see Sect. \ref{sec:full_mdot}). 

All models begin their evolution as a pre-MS object accreting material of uniform composition with H, He and metal content as follows. The Galactic and the LMC metal mass fractions are set to $Z = 0.02$ and $0.008$ respectively. The initial helium mass fraction $Y$ in our models is calculated as $Y = Y_{\text{prim}} + (\Delta Y/\Delta Z) Z$ where the primordial He abundance $Y_{\text{prim}} = 0.24$ and $(\Delta Y/\Delta Z) = 2$. The hydrogen mass fraction $X$ is calculated by the condition that the total mass fraction of all elements is 1, that is $X = 1-Y-Z$. Both  $Y_{\text{prim}}$ and $(\Delta Y/\Delta Z)$ follow the default values in MESA \citep{Pols1998}.

Convective mixing is treated using the standard mixing length theory (MLT) by \citet{MLT68}, parameterized by $\alpha_{\text{MLT}}$ expressed in terms of the pressure scale height $H_\mathrm{p}$. We choose a value of $\alpha_{\text{MLT}} = 1.5$ for all our models. The Ledoux convection criteria is used to define the convective boundaries. This allows for semiconvective regions to form above the convective core due to the chemical gradient formed by the receding core during the MS. A semiconvective diffusion efficiency of $\alpha_{\mathrm{sc}} = 1$ is used. 

Convective boundaries inside the star are set by the condition that the acceleration of the convective elements goes to zero. This does not necessarily mean these elements also have zero velocity at the core boundary. These elements can \textit{over-shoot} above (or below) the convective boundary and penetrate into the radiative regions with a non-zero velocity. Such overshooting above the convective regions can be described using different diffusion profiles. We use the exponential profile parameterized by $f_{\text{ov}}$ as described in \citet{Herwig2000}. The diffusion coefficient in the overshooting region is defined as follows:
\begin{equation}
\begin{array}{c@{\qquad}c}
D_{\text{ov}} = D_\text{o} e^{-2(r-r_\text{o})/f_{\text{ov}}H_\text{p}}
\end{array}
\label{exp_os}
\end{equation}
where $D_\text{o}$ is the diffusion coefficient at the start of the overshooting region defined at location $r_\text{o}$ just below the convective boundary, $f_{\text{ov}}$ defines the inverse slope of the diffusion decay expressed in terms of the pressure scale height $H_\mathrm{p}$. The exponential overshooting parameter $f_{\text{ov}}$ is approximately related to the step overshooting parameter $\alpha_{\text{ov}}$ as $f_{\text{ov}}\approx 0.1\alpha_{\text{ov}}$ \citep{Sabhahit2021}.

The amount of mixing above the convective core can significantly affect the evolution of massive stars, as more fuel is available for fusion. Higher overshooting mixing can change the MS lifetime of massive stars. A larger reservoir of fuel available takes longer to burn causing the MS lifetime to be extended. A larger core mass also leads to an increase in the surface luminosity as more H is converted to He. 

During the MS, thermal and hydrostatic equilibrium require the outer envelope layers of the star to expand to cooler temperatures while the convective core gradually contracts in size. Canonical O stars thus evolve redwards in the HR diagram during core H burning. As the core size increases with overshooting, the outer layers evolve to even cooler temperatures. Thus higher overshooting generally leads to more luminous and cooler stars at the end of the MS. But as we will see in Sect. \ref{sec:chem_hom}, overshooting has a negligible effect on the evolution of stars above a certain initial mass where the convective core encompasses almost the entire mass of the star and undergo chemically homogeneous evolution.

Closer to the surface, at cooler temperatures various opacity bumps are formed corresponding to the bound-free transitions of H, He and Fe. A convectively unstable layer is formed near such opacity peaks as $\nabla_{\mathrm{rad}} \sim \chi l/m$ can locally exceed the $\nabla_{\mathrm{ad}}$, where $l$, $m$ and $\chi$ are the local luminosity, mass coordinate and the total opacity and in general can vary with the radius $r$. 

Unlike the convection in the core, a small excess of the temperature gradient above the adiabatic temperature gradient is not sufficient to transport the entire energy of the star. Convection is not adiabatic and is termed `super-adiabatic' and both convection and radiation are responsible for transporting the energy. The fraction of the energy transported by convection decreases with increasing mass as radiation pressure begins to dominate the total pressure.

Using homologous stellar models composed of ideal gas, one can derive a simple relation between the luminosity and the stellar mass: $L \sim M^x,\; x\approx 3$. Such steep dependence of the luminosity on the stellar mass results in large $(L/M)$ ratios for VMS, and can bring them close to the so-called \textit{Eddington limit}. The Eddington limit defines the condition of outward radiative acceleration balancing the inward pointing gravitational acceleration. The `total' Eddington parameter that takes into account just the radiative luminosity $L_\mathrm{rad}$ is given by
\begin{equation}
\begin{array}{c@{\qquad}c}
\Gamma(r) = \dfrac{g_\mathrm{rad}}{g_\mathrm{newt}} = \dfrac{\chi(r)l_\mathrm{rad}(r)}{4\pi Gcm(r)}
\end{array}
\label{Gamma_tot_def}
\end{equation}
The total Eddington parameter can vary significantly with the radius $r$ through its dependence on the total local opacity $\chi(r)$ that includes both continuum and line sources. In the radiative regions of the star the entire luminosity is transported by the radiative diffusion: $l_\mathrm{rad}(r) = l(r)$, while in convective regions: $l_\mathrm{rad}(r) = l(r) - l_\mathrm{conv}(r)$. Near the surface, one can approximate the local luminosity $l(r)$ and mass $m(r)$ by the total luminosity $L$ and mass $M$ of the star. $G$ and $c$ are the gravitational constant and speed of light respectively

In contrast, the electron scattering opacity $\chi_{e}$ becomes nearly constant with radius for high enough temperatures where the H and He are fully ionized. The electron scattering Eddington parameter is given as
\begin{equation}
\begin{array}{c@{\qquad}c}
\Gamma_\mathrm{e} = \dfrac{\chi_{e}l}{4\pi Gcm} \approx 10^{-4.813}(1+X_\mathrm{s})\dfrac{L}{M}
\end{array}
\label{Gamma_e_def}
\end{equation}
where the right-most expression assumes the mean molecular weight per free electron $\mu_e \approx 2/(1+X_\mathrm{s})$ for fully ionized plasma where $X_\mathrm{s}$ is the surface hydrogen mass fraction, and the near-surface approximations for $l(r)$ and $m(r)$. The quantities $L$ and $M$ are in terms of the solar luminosity and solar mass respectively.

As the Eddington limit is approached $\Gamma(r)\rightarrow 1$, 1D models of massive stars that use standard mixing length theory predict the formation of very low density envelopes extended to very large radii where convection is highly inefficient \citep{Graf2012}. They find the inflation effect to largely depend on the topology of the iron opacity peak, essentially forming a density inversion near the surface to maintain hydro-static equilibrium. Massive stars with $M_\mathrm{init} \gtrsim 40 M_\odot$ can potentially exceed the Eddington limit locally already during the MS, corresponding to the ionization zones of iron (Fe), helium (He) and H, causing them to undergo inflation. The situation is worsened for cool supergiant stars that have access to the cooler H bump at log$(T_\mathrm{eff}) \approx 4$ where the Eddington parameter can reach values as high $\approx 7$ \citep{Sanyal2015}.

The physical existence of such statically inflated envelopes with density inversions is debated. Moreover, the near-sonic conditions in the super-adiabatic envelopes of massive stars is outside the applicability of the standard MLT theory of convection. Thus, specific routines exist in different evolutionary codes to remove such layers. In MESA, this is performed by a routine called MLT++ \citep{MESA13}. Radiative energy transport is decreased in favour of an alternate energy transport with potential candidates such as turbulent convection or radiation leakage through porous medium. By decreasing $L_\mathrm{rad}$, the Eddington parameter in Eq. \ref{Gamma_tot_def} can be suppressed below unity. The formulation of MLT++ thus allows it to suppress inflation and prevent the formation of density inversion in layers with low $\beta$ and high $\Gamma$. The different MLT++ parameters to gradually reduce the actual temperature gradient in these super-Eddington layers is described in detail in Appendix B of \citet{Sabhahit2021}. Recent studies have focused on the interplay between convective efficiencies in the envelopes and mass loss of core He burning supergiants and the associated Humphreys-Davidson limit \citep{Sabhahit2021, Agrawal2021}.

Rotation in MESA is treated using the shellular approximation, which allows for solving the inherently 3D process of rotation in a 1D approximation. Rotation can give rise to instabilities that can mix the interior of the star. The diffusion coefficients for both mixing and angular momentum transport corresponding to different rotation-induced instabilities follow \citet{Heger2000}. We ignore the effects of rotationally-enhanced mass loss in our models.

Rotation as a process can significantly affect the MS evolution of massive stars. Centrifugal forces arising from slow or moderate rotation can result in cooler stars due to lower surface effective gravity. However, rotation also induces internal mixing that can cause an opposing effect producing hotter and more luminous stars. Rapid rotation can also lead to chemically homogeneous mixing, where rotational mixing is strong enough to overcome the chemical gradient and completely mix the star from the center to its surface. All models presented here are solid-body rotators at the ZAMS with $\Omega/\Omega_{\text{crit}} = 0.2$. Akin to overshooting, rotation too has a negligible effect on the evolution of VMS \citep{Yusof2013, Kohler2015, Higgins2022}.  

At the highest masses, the temperature evolution of stellar models is highly sensitive to the input mass loss and the mixing in the envelope. As we show below, even small quantitative differences in the VMS mass loss can result in qualitatively different evolution of their surface properties. The observed effective temperatures of the VMS in the Arches and 30 Dor cluster are confined to a very narrow range of values, forming a band in the HR diagram.  In this study we intend to test VMS models with different mass loss dependencies on stellar
properties and investigate whether these models reproduce the narrow band of observed temperatures as a function of host metallicity.

Table \ref{table:input} summarizes the different inputs - the initial masses, the initial metallicities, the overshooting efficiency above the convective core and the rotational velocity expressed in terms of the critical velocity - all used to build our grid of models. The different mass-loss descriptions (V01, VMS - $\Gamma_\mathrm{e}$ and $L/M$) implemented in our models are detailed in Sec. \ref{sec:full_mdot}.

\begin{table}
\centering 

\begin{tabular}{c c} 

        \hline\hline
        $M_{\text{init}} (M_\odot)$ & 60, 80, 100, 120, 140, 160, 180, 200, 250, 300, 400, 500 \\
        $Z_{\text{init}}$ & 0.02, 0.008 \\
        $f_\text{ov}$ &  0.03  \\
        $\Omega/\Omega_{\text{crit}}$ & 0.2  \\
        $\dot{M}$ & V01, VMS ($\Gamma_\mathrm{e}$ and L/M) \\
        $\mathrm{MLT++}$ & on, off \\
        \hline
\end{tabular}
\caption{The initial masses, the initial metal mass fractions, the overshooting efficiencies and the rotation speeds used in our model grid } 
\label{table:input}
\end{table}

The mass fraction of metals in our models scale with the solar abundances taken from \citet{GS98}. The opacities used are from the OPAL Type 2 opacity tables, which takes into account of the increase in the mass fractions of metals as the star evolves. For the equation of state (EOS), we use the default tables available in MESA that are mainly based on the OPAL EOS tables. Outside the range covered by these tables, MESA switches to other EOS tables depending on the temperature and density. We refer to \citet{MESA11} for the different EOS options available. For the reaction rates, we use the \texttt{basic.net} network that includes 8 isotopes. Several complex networks are available in MESA especially those that are dedicated to implement reaction rates during the end stages of stellar evolution. Since our models only run until the end of the MS, we use the most `basic' reaction network available. 

\section{Main sequence mass loss}
\label{sec:mass_loss}

Theoretical efforts have been made to predict the mass loss as a function of stellar parameters for decades \citep{LS1970}. Analytical descriptions for the mass loss, mostly based on the CAK theory \citep{CAK1975} calculates radiative acceleration from lines assuming a simple power-law line strength distribution function. The CAK theory has been extended by relaxing the assumptions therein and accounting for complex line-lists and finite-disk effects and \citep{FA1986, Pauldrach1986, Kudritzki1989}. 

Theoretical mass-loss rates can also be derived by using the global energy Monte Carlo (MC) approach by \citet{AL1985}, where the lines are described in the Sobolev approximation. The fate of photospheric photons is tracked to calculate the radiative momentum transferred to the ions and the radiative acceleration $g_\mathrm{rad}(r)$. An advantage of MC approach is that it takes into account multiple scattering when evaluating $g_\mathrm{rad}$. \citet{MV2008} also get rid of the assumed $\beta$-law velocity stratification in the wind and solve the equation of motion self-consistently to simultaneously obtain mass-loss rates and velocity profiles. In recent years, a number of codes have the option to solve the radiative transfer in the co-moving frame (CMF) approach as well as to solve the full hydrodynamics to obtain both mass-loss rates and velocities \citep{Sander2017, KK2017, Bjorklund2021}. The CMF approach has the advantage of relaxing the Sobolev approximation which fails around the sonic point.

Both theoretical \citep{Vink2006, Vink2011} and empirical results \citep{Graf2011, Best2014} have hinted towards an enhanced mass loss for stars close to their Eddington limit. The challenge is to determine the mass-loss scaling with different stellar parameters and also the implement the correct absolute mass loss rates in stellar evolution codes. In Sec. \ref{sec:existing_vms} we present previous efforts in quantifying the mass loss of VMS, and the limitations of these implementations. In Sec. \ref{sec:VMS_mdot_full} we detail the results from \citet{Vink2001} and \citet{Vink2011} that are based on this MC approach to predict theoretical mass loss used in our models. We further discuss the so-called transition mass loss in Sec. \ref{sec:tmr}, a model-independent way to characterize the transition from optically-thin winds of O stars to the optically-thick winds of WNh stars at two different metallicities, and use it to calibrate our absolute mass loss (Sec. \ref{sec:full_mdot}).

\subsection{VMS mass loss implementations}
\label{sec:existing_vms}

Theoretical mass-loss rates for optically-thin winds of OB stars as a function of luminosity, mass, effective temperature, terminal velocity and host metallicity was provided by \citet{Vink2001}. A simple extrapolation of the canonical O-star rates have been used previously in the evolution of VMS \citep{Yusof2013, Kohler2015}. Here we provide the dependence of $\dot{M}$ on the luminosity $L$ and the mass $M$ of the star on the hot side of the bistability jump. The absolute rates are given in Eq. \ref{eq:optically_thin}.
\begin{equation}
\begin{array}{c@{\qquad}c}
\dot{M}_\mathrm{V01} \propto L^{2.194}M^{-1.313}
\end{array}
\label{low_gamma_mdot_depend}
\end{equation}
The positive exponent on the luminosity can be understood as the mass loss being stronger at higher luminosity owing to a larger wind driving force from the photons. The negative exponent on the mass is to do with the gravitational potential of the star. The heavier the star, the wind has to be driven from deeper down the potential well resulting in a lower mass loss.

The V01 rates quickly start to under-predict the mass loss above the transition point. This has been shown both theoretically \citep{Vink2006, Vink2011} and empirically in the 30 Dor cluster \citep{Best2014}. There have been efforts over the last decade to implement the correct mass loss beyond the V01 rates, for e.g. \citet{Yung2008} implement higher mass loss for stars close to the Eddington limit, but they adopt an ad-hoc equation for the VMS mass loss.

\citet{Graf2021} recently studied the evolution of VMS and suggested using a prescription that switches from an O-type wind to a Wolf-Rayet type wind on the basis of the parameter $\tau_\mathrm{s}$, the optical depth of the sonic point rather than employing the "Dutch" wind in MESA, which relies on a rather unphysical switch from O star to WNh regime based on a simple $X_\mathrm{s}$ condition. 

For models with $\tau_\mathrm{s}<2/3$, they use theoretical mass loss scaling predicted by \citet{Vink2001}, but with a $L$-scaling of 1.45 derived from empirical results in the 30 Dor \citep{Best2014}. This is shallower compared to the $L$-scaling of 2.194 given in Eq. \ref{low_gamma_mdot_depend}. However we note that the original mass-loss scaling predicted by \citet{Vink2001} varies with both luminosity and stellar mass. Using the mass-luminosity relation the mass loss-luminosity scaling reduces to $1.6$, in better agreement with the pure $\dot{M}-L$ scaling obtained empirically. Thus one has to careful when replacing the $L$-scaling of theoretical rates that vary with both luminosity and mass with a pure-$L$ empirical scaling where the stellar masses are unknown   \citep[see recent review by][]{VinkR}.

When the sonic point optical depth crosses unity, \citet{Graf2021} use an enhanced mass loss with a steep $\dot{M}-\Gamma_\mathrm{e}$ scaling of $5.22$ derived by fitting the empirical mass loss rates derived for WNh stars in the 30 Dor cluster \citep{Best2014}:
\begin{equation}
\begin{array}{c@{\qquad}c}
\mathrm{log}\dot{M}_\mathrm{Best2014}  = 5.22\; \mathrm{log}(\Gamma_\mathrm{e}) - 0.5\; \mathrm{log}(D) -2.6 
\end{array}
\label{eq:Graf_mdot}
\end{equation}
Mass-loss scaling with $\Gamma_\mathrm{e}$ can be obtained by assuming chemically homogeneous hydrogen and helium models to estimate the maximum and minimum mass for a star of given luminosity and surface hydrogen abundance. A similar steep $\mathrm{log} \dot{M}-\mathrm{log} \Gamma_\mathrm{e}$ slope of $4.293$ is obtained in the Arches cluster \citep{Graf2011}. The empirical analysis of VMS suggests a steeper dependence of the mass loss on the electron scattering Eddington parameter, and consequently a steeper implicit dependence on $X_\mathrm{s}$. 

However the assumptions made in the above analysis can drastically change the picture. First, the number of WNh star data points (13 and 8 in the Arches and 30 Dor respectively) available for the purpose of fitting are not too high compared to the fitting parameters involved, and one might run into the issue of over-fitting the data as pointed out in \citet{Graf2011}.

Secondly, the best-fit $\mathrm{log} \dot{M}-\mathrm{log} \Gamma_\mathrm{e}$ slope also depends significantly on the stellar masses used in the fitting process, with two extreme cases derived by assuming either chemically homogeneous models with given $X_\mathrm{s}$ or pure helium models. Assuming pure helium star masses, the obtained $\mathrm{log} \dot{M}-\mathrm{log} \Gamma_\mathrm{e}$ slope is $\approx -3$, which is qualitatively different from the slope of $\approx +4.2$ obtained for chemically homogeneous models. The true slope could be in between the two extremes cases considered here depending on the actual masses of these stars. This could change drastically the overall conclusions one can draw solely from such empirical fitting. 

Moreover the absolute rates derived from such empirical analysis also depend on the assumed clumping factor $D$ that characterizes the over-density of small-scale inhomogeneities in the wind compared to the inter-clump void regions. As we demonstrate in this paper, even tiny changes in the input mass loss of the order of 0.1 dex can significantly change the temperature evolution of VMS. Thus one must be cautious when using absolute mass loss rates derived empirically, that depends on the assumed clumping factor $D$, to determine mass loss recipes for stellar evolution. 

A different approach to implement VMS mass loss is discussed in \citet{Best2020mdot}, where they extend the O-star wind theory by \citep{CAK1975} (CAK) to accommodate the high-$\Gamma_\mathrm{e}$ regime of WNh stars:
\begin{equation}
\begin{split}
\mathrm{log}(\dot{M}_\mathrm{Best2020}) = &\;\;\mathrm{log}(\dot{M}_\mathrm{0}) + \Bigg(\dfrac{1}{\alpha} + 0.5\Bigg) \;\mathrm{log}(\Gamma_\mathrm{e}) \\ & - \Bigg( \dfrac{1-\alpha}{\alpha} +2\Bigg)\;\mathrm{log}(1-\Gamma_\mathrm{e})
\end{split}
\label{eq:best2020_mdot}
\end{equation}
where $\alpha$ is the CAK force multiplier parameter. The first term is the base mass loss rates. The second term proportional to $\Gamma_\mathrm{e}^{(1/\alpha)+ 0.5}$ dominates for low values of $\Gamma_\mathrm{e}$  and is relevant for canonical O stars, while the last term $(1-\Gamma_\mathrm{e})^{- ( \frac{1-\alpha}{\alpha} +2)}$ dominates in the high -$\Gamma_\mathrm{e}$ regime. This extended CAK formalism smoothly switches from one to the other where the two terms are equal. Equating the two terms, one can derive the values of $\Gamma_\mathrm{e}$ at the point of switch. Fitting the unclumped mass loss rates for the R136 stars, \citet{Best2020mdot} derived the value of $\alpha\approx 0.39$ which gives $\Gamma_\mathrm{e} = 0.47$ at the location of the switch (see also \citet{Brands2022}).

\begin{table}
\centering 

\begin{tabular}{c c} 

        \hline
        $\alpha$ & $\Gamma_\mathrm{e, switch}$ \\
        \hline\hline
        0 & 0.5 \\
        0.2 & 0.4849 \\
        0.4 & 0.4733 \\
        0.6 & 0.4641 \\
        0.8 & 0.45655 \\
        1 & 0.4503 \\
        \hline
\end{tabular}
\caption{The value of $\Gamma_\mathrm{e}$ for a wide range of CAK force parameter $\alpha$ values, at the location of switch from one mass loss dependency term to another in Eq. \ref{eq:best2020_mdot}. An solution can be obtained by equating the last two terms in the equation.} 
\label{table:gamma_e_alpha}
\end{table}

Although the recipe is simple and requires only one relation to quantify both the low and high-$\Gamma_\mathrm{e}$ mass loss, we note that the relation is foremost a purely mathematical one. The switch from one term to the other or the so-called `transition' is purely mathematical and may not represent the actual physical transition from canonical O stars to WNh stars. We test the sensitivity of this mathematical switch for different values of $\alpha$ in Table \ref{table:gamma_e_alpha}. We show that the $\Gamma_\mathrm{e}$ values at the switch is highly insensitive to the value of $\alpha$. For $\alpha$ values ranging from 0 to 1, the obtained values of $\Gamma_\mathrm{e}$ at the switch occupy a small range between 0.45 and 0.5. All other values outside this range is forbidden by such a relation and there is no reason for the exclusion of these values from a physical standpoint. The transition $\Gamma_\mathrm{e}$ values that we obtain in the two clusters even lies outside this range (see Table \ref{table:transition})

Moreover, empirical rates of WNh stars have been used to obtain the fits in the above analysis, which again is clumping-dependent. In this paper we use a clumping-independent method to calibrate the absolute mass loss rates of VMS, and thus the absolute rates are not affected by the uncertainties in the assumptions of inhomogeneities in the wind.

\subsection{$\Gamma_\mathrm{e}$ vs $L/M$-dependent mass loss}
\label{sec:VMS_mdot_full}

The absolute rates predicted by \citet{Vink2001} has been challenged in the mass range of canonical O stars, where there are uncertainties in the mass loss of a factor of 2-3. \citep{Bouret2003, Ft2006, Bjorklund2021}. These uncertainties are mostly due to the clumping in the wind and the treatment of radiative transfer in the sub-sonic region. However near the transition from O stars to WNh stars, the rates predicted by the MC simulations at Galactic metallicity closely match the model-independent transition mass-loss derived for the Arches Cluster \citep{Vink2012}. Thus just below the transition, concerning the lowest initial mass models in our grid (60 and 80 $M_\odot$), we use the absolute rates from \citet{Vink2001}.

\citet{Vink2011} performed MC calculations for stars with initial masses up to 300 $M_{\odot}$ and found a kink or upturn in the mass loss for the highest masses considered in their grid. The exact location of the kink could potentially be lower due to the limitations of self-consistently treating the deep sub-photospheric layers, and the under-prediction of the mass loss using the hydro-dynamically consistent treatment of the wind. 

Below the kink in the optically-thin O-star regime, they found their hydro-dynamically consistent results to agree well with the semi-empirical, $\beta$-velocity law approach from \citet{Vink2001}. Above the kink they found a very steep dependence of the mass loss on $\Gamma_\mathrm{e}$ as was already predicted in \citet{Vink2006}. They also found a change in the spectral morphology of the He\,\textsc{ii} line at 4686 \AA, a characteristic spectral line of Of and WNh stars as they crossed the transition point, suggesting a natural transition from normal O stars to the intermediate Of/WN stars and finally the WNh sequence. 

\citet{Vink2011} do not provide absolute mass-loss rates as the new dynamical approach tends to over-predict the terminal velocities resulting in under-predicted rates, but they provide the dependencies of the mass loss on stellar parameters such as $\Gamma_\mathrm{e}$ and $M$, 
\begin{equation}
\begin{array}{c@{\qquad}c}
\dot{M}_\mathrm{\Gamma_{e}} \propto M^{0.78}\Gamma_\mathrm{e}^{4.77} \propto \dfrac{(1+X_\mathrm{s})^{4.77}L^{4.77}}{M^{3.99}}
\end{array}
\label{high_gamma_mdot_depend}
\end{equation}
which we henceforth refer to as the $\Gamma_\mathrm{e}$-dependent VMS wind. 

We emphasize on two key points regarding the mass loss scaling in Eq. \ref{high_gamma_mdot_depend}. First, the theoretical mass-loss rates predicted by the MC method could not be fit by a power law of just one parameter. That is, the mass loss predicted varied over a range of values even for a fixed value of $\Gamma_\mathrm{e}$. Unless one assumes a specified mass-luminosity relation, the theoretical rates predicted by both \citet{Vink2001} and \citet{Vink2011} are a function of either $L$ and $M$, $L$ and $\Gamma_\mathrm{e}$, or $M$ and $\Gamma_\mathrm{e}$. The relation of stellar mass and its luminosity can vary with the internal structure and the surface H abundance, and is hard to predict unless chemical homogeneity is assumed.

The second important point is regarding the dependency of the mass loss on $X_\mathrm{s}$. In the first instance, \citet{Vink2011} computed a large grid of VMS models for a fixed value of $X_\mathrm{s} = 0.7$. The obtained $\dot{M}-\Gamma_\mathrm{e}$ slope is the same as the $\dot{M}-(L/M)$ slope. They later increase the He abundance in selected models for fixed values of $L$ and $\Gamma_\mathrm{e}$  by decreasing the mass, and find an increase in the mass loss, while still retaining the kink behavior in the mass loss. MC and CMF models have so-far shown a weak dependence of mass loss on the H abundance. This has generally been interpreted in the sense that while the H and He opacity change the underlying SED, the driving of metals dominates, and replacing H for He does not make much change to the line driving \citep{VdK2002, GH2008, Vink2011}. However some of these dependencies \citep{GH2008, Vink2011} have been computed for fixed $\Gamma_\mathrm{e}$ and the additional specific dependence on the number of electrons has not yet been established.

The over-predicted terminal velocities obtained from the MC models compared to the empirical values might hinder the determination of the exact ($1+X_\mathrm{s}$) scaling of mass loss. While there is evidence from both theory and observations that stars near the Eddington limit, i.e.,  high $L/M$, have increased mass loss in comparison to canonical O stars \citep{Vink2011, Graf2011, Best2014}, the exact implementation is still unknown. 


In this paper we study the VMS evolution of two extreme scenarios, based on the dependence of mass loss on the surface hydrogen. We compare the VMS models with a steep $\Gamma_\mathrm{e}$-dependence in Eq. \ref{high_gamma_mdot_depend} with an extreme case of a pure $L/M$-dependent wind where the $1+X_\mathrm{s}$ term is dropped entirely, a trend similar to the mass loss rates of canonical O stars (cf. eqn. 4):
\begin{equation}
\begin{array}{c@{\qquad}c}
\dot{M}_\mathrm{L/M} \propto M^{0.78}(L/M)^{4.77} \propto \dfrac{L^{4.77}}{M^{3.99}}
\end{array}
\label{high_gamma_LM}
\end{equation}
which we refer to as the $L/M$-dependent VMS wind.  While we test an extreme scenario of no hydrogen dependence in our models here, we are not implying that $1+X_\mathrm{s}$ has no effect on $\dot{M}$ but only that it is smaller compared to the $L/M$ effect. Perhaps a better understanding of mass-loss dependencies can be achieved through state-of-the-art stellar atmosphere codes such as PoWR that can consistently solve the hydro-dynamics of the wind stratification coupled with statistical equations and radiative transfer in the co-moving frame (CMF).

\citet{Vink2011} also performed additional simulations by varying the effective temperature to study its effect on the mass loss. For the models above the kink they find the mass loss to only weakly depend on the surface temperature (see their Fig. 7). As we will see below, our VMS models evolve at almost constant temperatures (see Fig. \ref{fig:HRD_gal}) during the main sequence.  A temperature dependency in the input mass loss should  thus not change our conclusions.

To summarize, there is theoretical and observational evidence to support an increase in the mass loss for stars close to the Eddington limit. Both agree well on the steeper $L$-dependence of the mass loss for the optically-thick  winds of WNh stars. However MC calculations do not find evidence of a similar strong dependence on the surface H, and might hint towards a weaker scaling compared to the empirical results in the two clusters. For this reason, we study the MS evolution of VMS using the two dependencies in Eq. \ref{eq:optically_thick_gamma} and \ref{eq:optically_thick_LM} to highlight the importance of mass-loss scaling with the surface H and how it can cause qualitative differences in the evolution of VMS. We directly compare and contrast VMS models that use different dependencies and compare their temperature evolution in Sect. \ref{sec:results}.

\begin{table*}
\centering 

\begin{tabular}{c c c c c c c c} 
        \hline

         & $L/L_\odot$ & $\varv_\infty (\mathrm{km/s})$ & $T_{\mathrm{eff}}\mathrm{(K)}$ & $X_{s}$ & $\dot{M}_{\mathrm{trans}}$ & $M/M_\odot$ & $\Gamma_{\mathrm{e},{\mathrm{trans}}}$ \\
        \hline\hline
        GAL & $10^{6.06}$ & 2000  & $33900$  &  $0.7$ & $-5.16$ & $76.9$ & 0.39 \\
        LMC &  $10^{6.31}$ & 2550  & $44400$  &  $0.62$ & $-5.0$ & $121.8$ & 0.42 \\

        \hline
\end{tabular}
\caption{Transition parameters of Of/WNh stars from Arches cluster in the Galaxy and 30 Dor cluster in the LMC. The transition luminosities, effective temperatures and $X_\mathrm{s}$ are obtained empirically, and used to evaluate the transition mass-loss rate and the $\Gamma_{\mathrm{e},{\mathrm{trans}}}$ } 
\label{table:transition}
\end{table*}

\subsection{Transition mass loss}
\label{sec:tmr}

\citet{Vink2012} derive a model-independent way to characterize the physical transition from optically-thin winds of O stars to the optically-thick winds of WNh stars, regardless of the assumptions of clumping in the wind. They argue that a simple relation between the sonic optical depth parameter $\tau$ and the wind efficiency parameter $\eta$ can be derived at the transition point under the assumption of very high $\Gamma(r)$, the total Eddington parameter. This condition $\eta = \tau  =  1$ can be used  to obtain a model-independent mass loss, called the transition mass loss. 

The $\eta = \tau  =  1$ condition at the transition relies on the assumption that the ratio $(\Gamma-1)/\Gamma$ is very close to unity or $\Gamma \gg 1$. \citet{Vink2012} test this assumption of high $\Gamma$ using a hydro-dynamic model of WR22 from \citet{GH2008}, by numerically integrating inwards from infinity to the sonic radius to obtain $\tau$. They derive a correction factor $f$ for their transition condition: $\eta = f\tau$. They further run a simple $\beta$ law model and find very similar correction factor $f$ as the hydro-dynamic model (see their Table 1).  A correction factor $f \simeq 0.6$ is obtained when the observed $\varv_{\infty}$ is used in their calculations. This gives the transition mass loss as follows
\begin{equation}
\begin{array}{c@{\qquad}c}
\dot{M}_{\mathrm{trans}} = f\dfrac{L_{\mathrm{trans}}}{\varv_{\infty} c}
\end{array}
\label{transition_theory}
\end{equation}
where $\dot{M}_{\mathrm{trans}}$ can be calculated given the luminosity and the terminal velocity of the Of/WNh transition stars. In fact, \citet{Vink2012} evaluate a transition mass-loss rate of $\mathrm{log}(\dot{M}_{\mathrm{trans}}) = -5.2$ in the Arches cluster using this formula. This value agrees well with the empirical mass loss derived for the Of/WNh-transition-type objects in the Arches cluster with a clumping factor $D = 10$ \citep{Martins2008}. 

With an average $L_{\mathrm{trans}} = 10^{6.06}L_\odot$, $\varv_{\infty} = 2000\; \mathrm{km/s}$ and $f = 0.6$ for the two transition stars in the Arches Cluster: F15 and F10, we obtain a $\mathrm{log}(\dot{M}_{\mathrm{trans}}) = -5.16$. \citet{Vink2012} also found that the mass loss of the transition objects in the Arches cluster agrees with the theoretical rates predicted from \citet{Vink2001} for Galactic metallicity. Assuming the V01 rates correctly predict the mass loss of these transition stars, one can derive an `average' mass of $M/M_\odot = 76.9$. Combined with $L_{\mathrm{trans}}$ and $X_\mathrm{s,trans}$ one can obtain the $\Gamma_{\mathrm{e},{\mathrm{trans}}}$, the location of the kink or upturn in the mass loss. This provides a model-independent transition mass loss where the winds switch from optically-thin to the optically-thick regime in the Arches cluster.

One can perform a similar analysis for the 30 Dor cluster in the Large Magellanic Cloud where we have empirical results of transition stars \citep{Best2014}. There are six stars with the Of/WN subtype, whose luminosities and terminal velocities can be averaged over to obtain the $\dot{M}_{\mathrm{trans}}$. We make an assumption that the correction factor $f$ is independent of metallicity and take the value $f = 0.6$ derived previously. This gives a $\mathrm{log}(\dot{M}_{\mathrm{trans}}) = -5.0$, which agrees well with the empirical results we have in the LMC again for a clumping factor of $D = 10$.

The stellar parameters of the transition stars in the two galaxies are summarized in Table \ref{table:transition}. We use these parameters to derive a mass loss recipe that connects the low-$\Gamma_\mathrm{e}$, optically-thin mass loss predicted by \citet{Vink2001}, with the high-$\Gamma_\mathrm{e}$ optically-thick mass-loss dependencies of \citet{Vink2011}.

\subsection{The complete mass loss recipe}
\label{sec:full_mdot}

We set the derived transition mass loss in the respective galaxies as the condition to switch from an optically-thin O-star wind to a wind that is dense enough to begin optically-thick wind driving. This condition is motivated from the wind efficiency parameter $\eta$ surpassing the single scattering limit above the transition point as found in \citet{Vink2011}. 

For the low-$\Gamma_\mathrm{e}$ regime, we use the standard O-star rates from \citet{Vink2001}. The \textbf{V01 rates} for the hot side of the bistability jump ($T_{\mathrm{eff}} > 25\; \mathrm{kK}$) is given as:
\begin{equation}
\begin{split}
\mathrm{log} \;\dot{M}_\mathrm{low, V01} = \;\;& -6.697 + 2.194\;\mathrm{log}(L/L_\odot/10^5) \\ & -1.313\;\mathrm{log}(M/M_\odot/30)\\ & -1.226\;\mathrm{log}((\varv_\infty/\varv_{\mathrm{esc}})/2) \\ & +0.933\; \mathrm{log}(T_{\mathrm{eff}}/40000) \\ & -10.92\;\{\mathrm{log}(T_{\mathrm{eff}}/40000)\}^2 \\
&  + 0.85 \;\mathrm{log}(Z/Z_\odot)
\end{split}
\label{eq:optically_thin}
\end{equation}
where the terminal velocity $\varv_\infty$ is a power law function of the metallicity according to \citet{Leitherer1992}, $\varv_\infty \sim Z^{0.13}$.

The V01 rates were derived for optically-thin winds of O stars with $\Gamma_\mathrm{e} \lesssim 0.4$. It includes a $\dot{M}-Z$ dependence for $1/30 < Z/Z_\odot < 3$, with the mass loss increasing with the host metallicity due to more metal lines being available for the wind driving. This dependence on the surface $Z$ is actually a proxy for the surface Fe  abundance, as iron is the dominant wind driver in the region below the sonic point where the  mass loss is fixed \citep{Vink99}. Accordingly we implement a $\dot{M}-Z$ dependence that only reflects changes in surface Fe abundance as originally intended in \citet{Vink2001}.  

For optically-thick winds of WNh stars above the transition mass loss, we provide both the $\Gamma_\mathrm{e}$-dependent and the $L/M$-dependent mass loss recipe, the former having a strong ($1+X_\mathrm{s}$) dependence and the latter having no dependence on the surface H. As mentioned previously, \citet{Vink2011} predict only a weak dependence of the high-$\Gamma_\mathrm{e}$ mass loss on the effective temperature. Considering the two unknowns to be the base mass-loss rate $\dot{M}_\mathrm{o}$ and the metallicity dependence $f_z$, and given the transition points at two different metallicities, we can evaluate both these values using the  following constraint
\begin{equation}
\begin{split}
\dot{M}_\mathrm{low} =  \dot{M}_\mathrm{trans} =  \dot{M}_\mathrm{high}
\end{split}
\label{eq:equality_condition}
\end{equation}
We obtain $\dot{M}_\mathrm{o} = -9.552$ and $f_z = 0.5$ for the case of $\mathbf{\Gamma_\mathrm{e}}$\textbf{-dependent mass loss}: 
\begin{equation}
\begin{split}
\mathrm{log}\; \dot{M}_\mathrm{high,\; \Gamma_\mathrm{e}} = \;\;& -9.552 + 4.77\;\mathrm{log}(1+X_\mathrm{s}) \\ & + 4.77\;\mathrm{log}(L/L_\odot/10^5)\\ & -3.99\;\mathrm{log}(M/M_\odot/30)\\ & -1.226\;\mathrm{log}((\varv_\infty/\varv_{\mathrm{esc}})/2) \\ 
&  + 0.5 \;\mathrm{log}(Z/Z_\odot)
\end{split}
\label{eq:optically_thick_gamma}
\end{equation}

As mentioned earlier, we want to test the effects of having a weaker ($1+X_\mathrm{s}$) dependence on the evolution of VMS. Thus we perform the above procedure yet again, but with no surface H dependence which gives the coefficients for the $
\mathbf{L/M}$\textbf{-dependent mass loss} as $\dot{M}_\mathrm{o} = -8.445$ and $f_z = 0.761$.
\begin{equation}
\begin{split}
\mathrm{log}\; \dot{M}_\mathrm{high,\; L/M} = \;\;& -8.445 + 4.77\;\mathrm{log}(L/L_\odot/10^5)\\ & -3.99\;\mathrm{log}(M/M_\odot/30)\\ & -1.226\;\mathrm{log}((\varv_\infty/\varv_{\mathrm{esc}})/2) \\ 
&  + 0.761 \;\mathrm{log}(Z/Z_\odot)
\end{split}
\label{eq:optically_thick_LM}
\end{equation}
Both $\Gamma_\mathrm{e}$ and $L/M$-dependent mass loss recipes imply higher mass loss than canonical O-star rates. As long as stars remain hot and ionized the electron number density remains constant. But if He reaches the surface through mixing, the number of electrons can change — and the two recipes begin to diverge. The absolute mass loss predicted by the two recipes differ by an order of 0.1 dex, the typical error bar on the empirical mass loss estimate of massive stars.

The obtained $Z$-dependence of the mass loss driven by optically-thick winds is shallower compared to that of optically-thin winds. This is in agreement with the theory as continuum processes begin dominating over the line process in driving the wind. Note however that this $Z$ dependence is derived by considering the properties at the transition point, and could indeed be shallower for stars well above the transition point.

\section{MESA evolution results}
\label{sec:results}

The VMS models we present here smoothly switch from the low and high-$\Gamma_\mathrm{e}$ mass loss by choosing the maximum of the two. Such an implementation automatically uses the low-$\Gamma_\mathrm{e}$ recipe from Eqn. \ref{eq:optically_thin} below the transition and the high-$\Gamma_\mathrm{e}$ recipe from Eqn. \ref{eq:optically_thick_gamma} or \ref{eq:optically_thick_LM} above it. As detailed in Sec. \ref{sec:tmr}, such a transition can be calculated by a model-independent method based on a simple condition connecting the sonic optical depth and the wind efficiency parameter $\eta = f\tau$. 

We begin by comparing the temperature evolution of VMS test models at Galactic and LMC metallicities to the empirical results from the VMS in Arches and 30 Dor cluster respectively. To this end, we run a small number of test models with MLT++ off, and use three different mass-loss dependencies as detailed in Sec. \ref{sec:temp_evolution_VMS}. We further explore models where inflation is suppressed and discuss the possible degeneracy in the temperature evolution of VMS (Sect. \ref{sec:MLT++}). We then provide a full grid of VMS models at Galactic metallicity that implement our $(L/M)$-dependent mass loss and compare them with empirical results available in the Arches cluster (\ref{sec:model_grid}). We also explore the chemically homogeneous evolution of VMS above a certain initial mass (\ref{sec:chem_hom}). Furthermore, we compare the Galactic models with the LMC models to study the role of metallicity in VMS evolution in Sect. \ref{sec:metallicity_effect}.

\subsection{Temperature evolution of VMS}
\label{sec:temp_evolution_VMS}

The VMS in the Arches cluster all have very similar temperatures and form a narrow band in the HR diagram with log$(T_\mathrm{eff}) \approx 4.5-4.6$ above log$(L/L_\odot) \approx 6.0$. The cluster is located close to the Galactic center and the $Z$ content might be super-solar  \citep{Martins2008}. \citet{Crowther2010} analysed the WN5-6h stars in the two star clusters NGC3603 and R136 in the Galaxy and LMC respectively. The temperatures in NGC3603 ($\sim$ 4.62) were systematically lower by $\approx 0.1$ dex than the VMS in R136 ($\sim$ 4.72). This feature is also observed in the 30 Dor cluster with temperatures between log$(T_\mathrm{eff}) \approx 4.6-4.7$ \citep{Best2014}. Although there is an overlap in the VMS temperatures of NGC3603 and the 30 Dor, it is well within the error margin for the above correlation to hold.

\begin{figure}
    \includegraphics[width = \columnwidth]{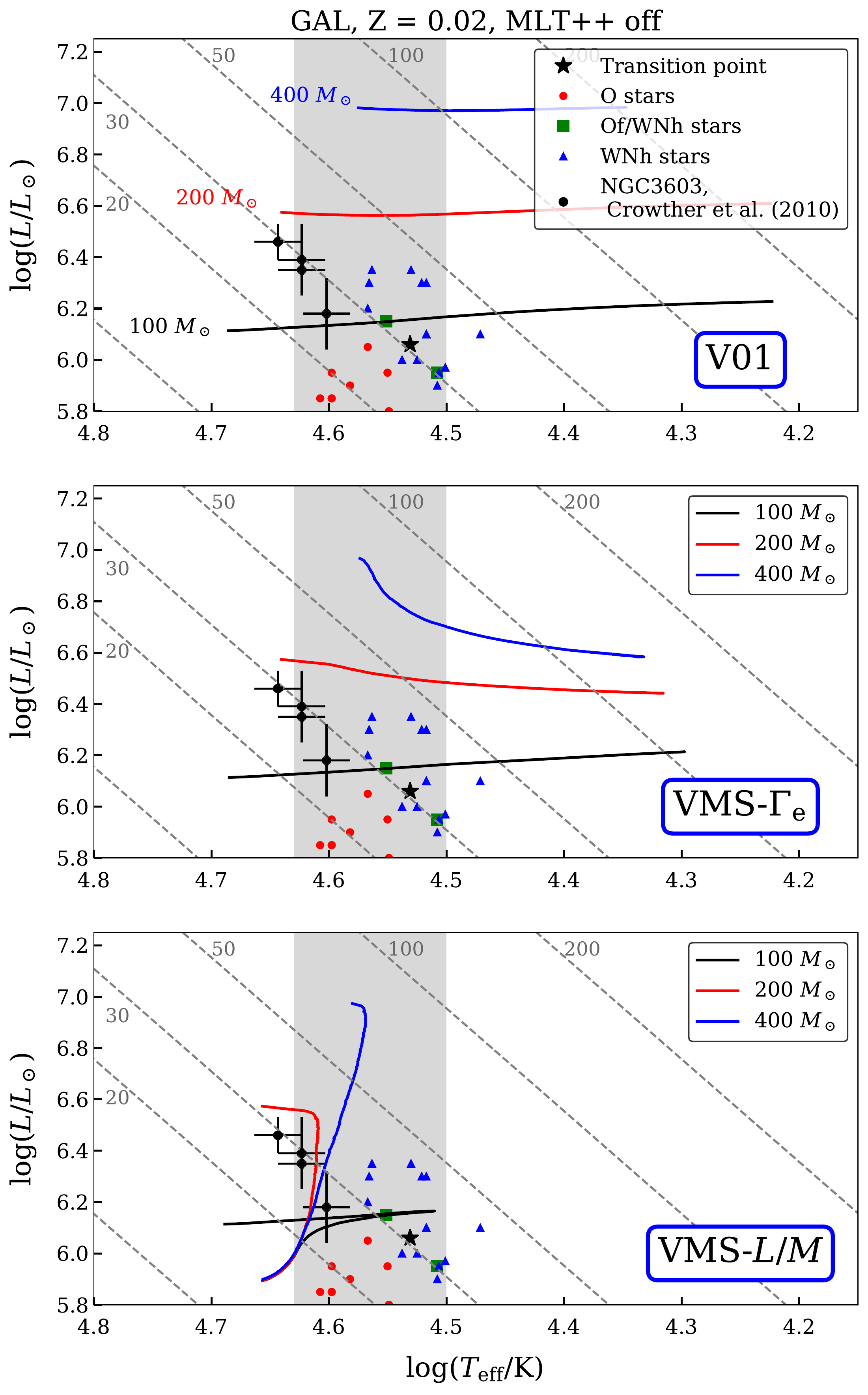}
    \caption{Stellar tracks of VMS test models at Galactic metallicity and MLT++ off. The initial masses considered at 100, 200 and 400 $M_\odot$. The three rows use three different dependencies on the mass loss as described in the text. Empirical luminosities and effective temperatures of VMS in the Arches cluster \citep{Martins2008} are shown in colored symbols representing different classes of stars. The `average' surface properties of the transition stars used in the analysis are represented by the black star symbols. The four WN6h stars analysed by \citet{Crowther2010} in the NGC3603 cluster are shown in black circles.} The grey dashed lines are constant radius lines and the numbers represent the radius in terms of $R_\odot$.
    \label{fig:HRD_gal_TEST}
\end{figure}

\begin{figure}
    \includegraphics[width = \columnwidth]{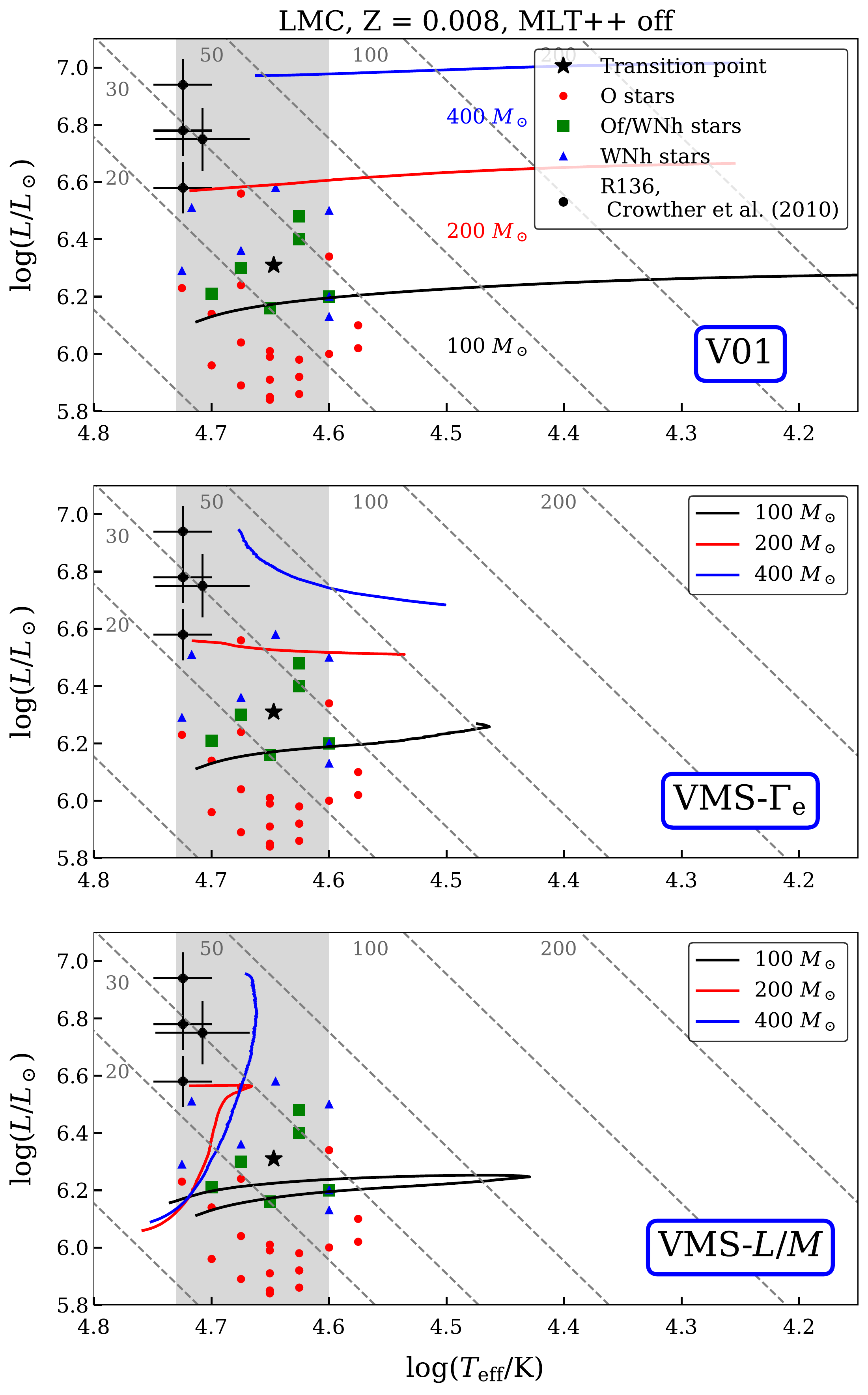}
    \caption{Stellar tracks with same input physics as as Fig. \ref{fig:HRD_gal_TEST} but at a lower metallicity corresponding to the LMC. The empirical results from 30 Dor \citep{Best2014} are shown in colored symbols (following the same notation as Fig. \ref{fig:HRD_gal_TEST}). The brightest four objects from the central R136 region \citep{Crowther2010} are also plotted (black circles).}
    \label{fig:HRD_lmc_TEST}
\end{figure}

In Fig. \ref{fig:HRD_gal_TEST} we plot the empirical luminosities and effective temperatures of the VMS in the Arches cluster (colored symbols for different classes of stars: O - red circles, Of/WNh - green squares and WNh stars - blue triangles) \citep{Martins2008} near the Galactic center and the four WN6h stars in the NGC3603 cluster (black circles) from \citet{Crowther2010}. In the LMC (Fig. \ref{fig:HRD_lmc_TEST}), we take the empirical results from \citet{Best2014} who analysed the VMS in the 30 Dor cluster, and also plot the brightest four objects (black circles) from the central R136 region from \citep{Crowther2010}. The `average' surface properties of the transition stars used as the anchor point to calibrate the mass loss is represented by the 'black star' symbol. 

We compare the empirical results with stellar tracks with $Z = 0.02$ and $0.008$ and use the $\Gamma_\mathrm{e}$ and $L/M$-dependent mass loss.
These test models are allowed to inflate by switching off MLT++ mixing in the envelope (for models with MLT++ on, see Sec. \ref{sec:MLT++}). The initial masses considered here for the comparison are 100, 200 and 400 $M_\odot$ (black, red and blue lines) and are evolved till core hydrogen exhaustion (for the full grid see Fig. \ref{fig:HRD_gal} and \ref{fig:HRD_lmc}). All inputs are fixed and we only change the mass-loss scaling above the transition used during the evolution. We evolve the test models with an overshooting mixing efficiency of $f_\text{ov} = 0.03$ and rotation of $\Omega/\Omega_{\text{crit}} = 0.2$. The case of different rotational and overshooting input parameters is mentioned in Sec. \ref{sec:chem_hom}.

We compare three different mass loss descriptions listed below.
\begin{enumerate}
  \item The first set of models in the top row use the optically-thin wind mass loss given in Eq. \ref{eq:optically_thin}, and are called the V01 models. Although the initial mass of the star is greater than the transition mass in the Galaxy (Table \ref{table:transition}) and the V01 rates are known to severely under-predict the mass loss in this mass range, we run these test models for the purpose of comparison.
  \item The second set in the middle row use $\Gamma_\mathrm{e}$- dependent mass-loss rate from Eq. \ref{eq:optically_thick_gamma} that includes a strong $1+X_\mathrm{s}$ dependence. We note that stars with initial mass near the transition point can switch from the optically-thin to optically-thick regime during their evolution. The 400 $M_\odot$ model is well above the transition and uses on the optically-thick $\Gamma_\mathrm{e}$- dependent mass loss throughout its evolution.
  \item And for the bottom row, we use the $L/M$-dependent mass loss from Eq. \ref{eq:optically_thick_LM} with no $1+X_\mathrm{s}$ dependence and a pure $L/M$-dependence. The absolute rates of both second and third test models are calibrated by the transition mass loss.
\end{enumerate}

The models in the top row (V01) quickly evolve redwards, as both increased mass loss and efficient MLT++ mixing are absent to suppress the radius inflation effect as these stars are very close to the Eddington limit. These stars also quickly inflate outside the range of the observed temperatures in both the Galaxy and the LMC (but see Sec. \ref{sec:MLT++}) and can even cross the HD limit.

The middle row models use the steep $\dot{M}-\Gamma_\mathrm{e}$ scaling and overall higher absolute mass loss compared to the top row. The higher mass-loss rate keeps the effect of inflation under control, in agreement with the increased $\Gamma_\mathrm{e}$-dependent mass loss models of \citet{Graf2021}. When compared to the fully inflated models from top row the tracks still evolve almost horizontally but are more hotter and compact.

\citet{Graf2021} present a grid of VMS models at LMC metallicity to explain the narrow width of the effective temperatures of the observed VMS in the 30 Dor cluster. They argue that at the highest masses, models with MLT++ on evolve bluewards and flip over to hotter temperatures, and might not might not reflect what VMS do in Nature as the empirical results of VMS in the 30 Dor cluster suggest much cooler and narrower range of temperatures. They suggest that while tthe effect of inflation can cause the models at the highest masses to evolve to cooler temperatures, using a steeper $\dot{M}-\Gamma_\mathrm{e}$ mass loss with absolute rates derived empirically from \citet{Best2014} might suppress this inflation and potentially explain the narrow temperature feature. 

However, we note that at these highest masses and luminosities, there is a delicate balance between the effects of increased mass loss and inflation. The inflation and increased mass loss have opposing and mass  dependent effects on the temperature of VMS. Inflation can cause the highest masses to evolve to cooler temperatures. Mass loss is also stronger at higher masses and can cause the star to evolve to hotter temperatures by quickly stripping the envelope and exposing the deeper, hotter layers. 

As long as VMS evolve horizontally at high luminosities, both effects significantly affect their evolution. While it may be possible to have constant temperatures when stars evolve horizontally as \citet{Graf2021} demonstrate for the VMS in the 30 Dor, but it is quite challenging to exactly balance the two effects: inflation and mass loss — over the entire mass range and across different $Z$. The VMS temperatures are thus highly sensitive to the input mass loss and the energy transport efficiencies in the envelope.



Stars in the bottom row evolve at almost constant temperatures and drop in luminosity. The evolution is qualitatively different despite having typically only $0.1-0.2$ dex difference in the absolute rates compared to the models in the middle row. This drop in the luminosity has a self-regulatory action in maintaining constant temperatures throughout the evolution. The inflation effect reduces with decreasing luminosity potentially causing the temperatures to flip over to the hot side, but so does the mass loss, preventing thereby a complete shift in the temperatures to the hotter side. Vertical evolution would give a natural way to account for constant temperatures over entire mass range, as well as explain the temperature trend with $Z$. The slightly hotter temperatures predicted by the model might indicate a super-solar metallicity for the stars in the Arches cluster as was already hinted in \citet{Martins2008}.

\begin{figure}
    \includegraphics[width = \columnwidth]{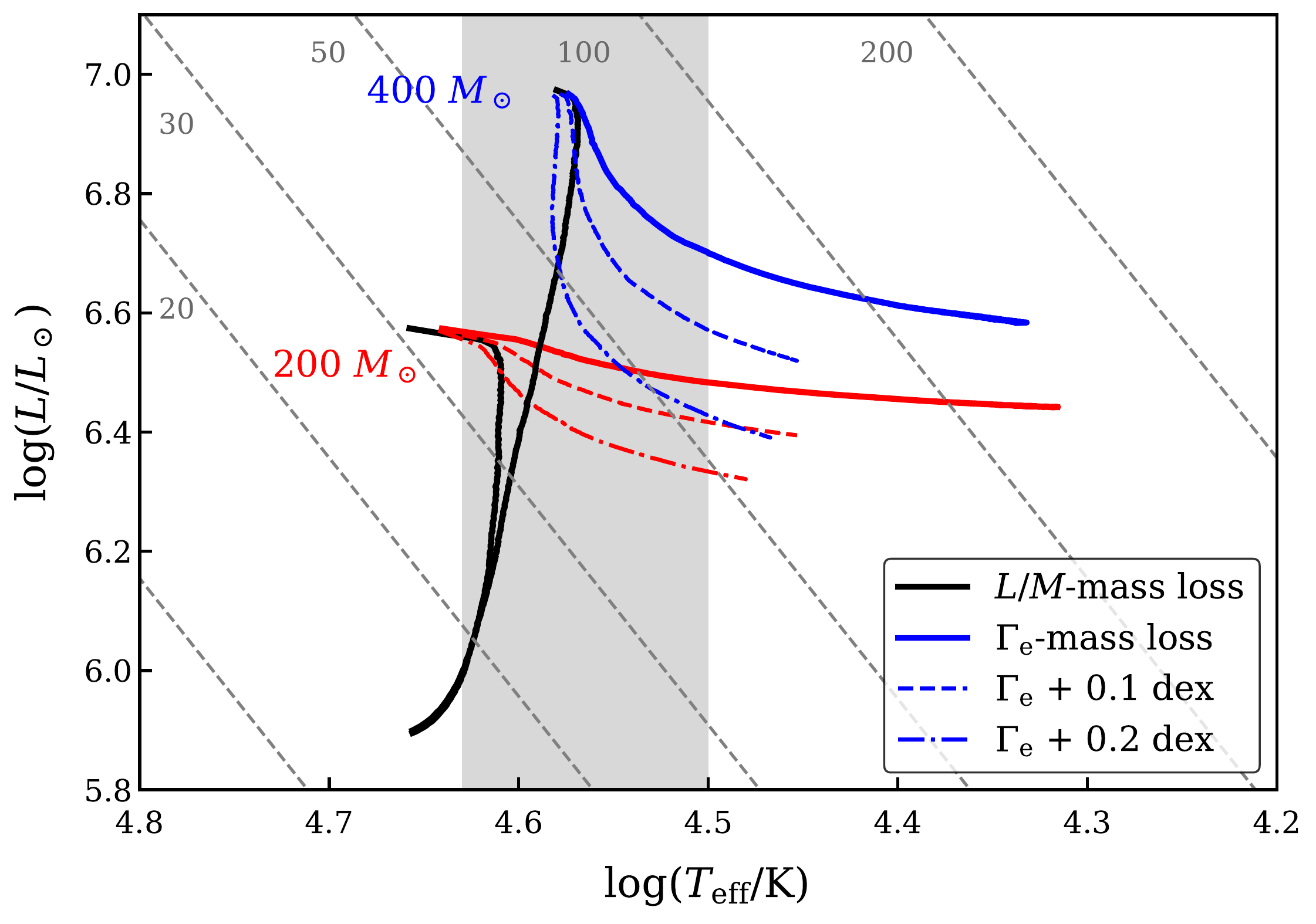}
    \caption{Stellar tracks of 200 (red) and 400 $M_\odot$ (blue) Galactic models where the absolute rates of the $\Gamma_\mathrm{e}$-dependent mass loss are artificially boosted by 0.1 (color dashed) and 0.2 dex (colour dot-dashed) at ZAMS. We also plot the original $\Gamma_\mathrm{e}$ (colour solid) and $L/M$-dependent mass loss (black solid) models from Fig. \ref{fig:HRD_gal_TEST}. }
    \label{fig:hrd_absolute}
\end{figure}

To show that the temperature evolution of the middle and bottom row models are indeed qualitatively different, we run additional test models where the absolute rates of the $\Gamma_\mathrm{e}$-dependent mass loss are artificially boosted by 0.1 and 0.2 dex at ZAMS, and compare them with the $L/M$-dependent mass loss which is typically $0.1-0.2$ dex higher. In Fig. \ref{fig:hrd_absolute}, we see that despite artificially increasing the absolute rates of the $\Gamma_\mathrm{e}$-dependent mass loss by 0.2 dex at ZAMS, they still evolve qualitatively different compared to the $L/M$-dependent mass loss. This is due to the strong dependence on the surface hydrogen, that can severely suppress the mass loss as the star evolves. While the horizontal evolution is halted by the increase in absolute rates by $0.1-0.2$ dex, we note that the increase in the absolute rates required could be arbitrary and vary with the initial mass or metallicity.

To summarize, the temperatures of VMS seem to correlate with the metallicity $Z$, with cooler temperatures at higher $Z$. The tracks evolving vertically can naturally explain the narrow temperature feature, through a self-regulatory mechanism of decreasing luminosity that suppresses both inflation and mass loss, giving a robust temperature independence in the evolution of VMS. It would also be premature to exclude the case of horizontal evolution using the $\Gamma_\mathrm{e}$-scaling as the stars of the same age $~0-0.5$ Myr would have very similar temperatures. However, we also note that the tight temperature-$Z$ correlation might hint towards a dominant $Z$ effect, which would be easier to explain with vertical evolution despite the different cluster ages.

\subsection{Effect of efficient envelope mixing}
\label{sec:MLT++}

As described earlier, MLT++ suppresses inflation by reducing the temperature gradients closer to the surface. This can change the radius evolution of massive star models, even changing the location of the  ZAMS at the highest masses. When switched on, this routine acts on the super-adiabatic layers in the opacity bumps. 

In this section we explore the possibility of a degeneracy in the temperature evolution of VMS. In the previous section, we showed that an increased mass loss closer to the Eddington limit given by a pure $L/M$-dependence can potentially explain the observed temperatures of VMS. However a very similar qualitative behavior of constant temperature evolution could also be reproduced instead by increasing the mixing in the envelopes of these stars thus `balancing' the inflation. 

To test how the convective mixing treatment in the envelopes of VMS affects their evolution, we run two test models, with initial mass of 100 and 400 $M_\odot$ that only differ from our previous calculations in the energy transport efficiency in their envelope:
\begin{enumerate}
    \item MLT++ on and low-$\Gamma_\mathrm{e}$ mass loss from Eq. \ref{eq:optically_thin}.
    \item  MLT++ on and high-$\Gamma_\mathrm{e}$ mass loss from Eq. \ref{eq:optically_thick_LM}.
\end{enumerate}

\begin{figure}
    \includegraphics[width = \columnwidth]{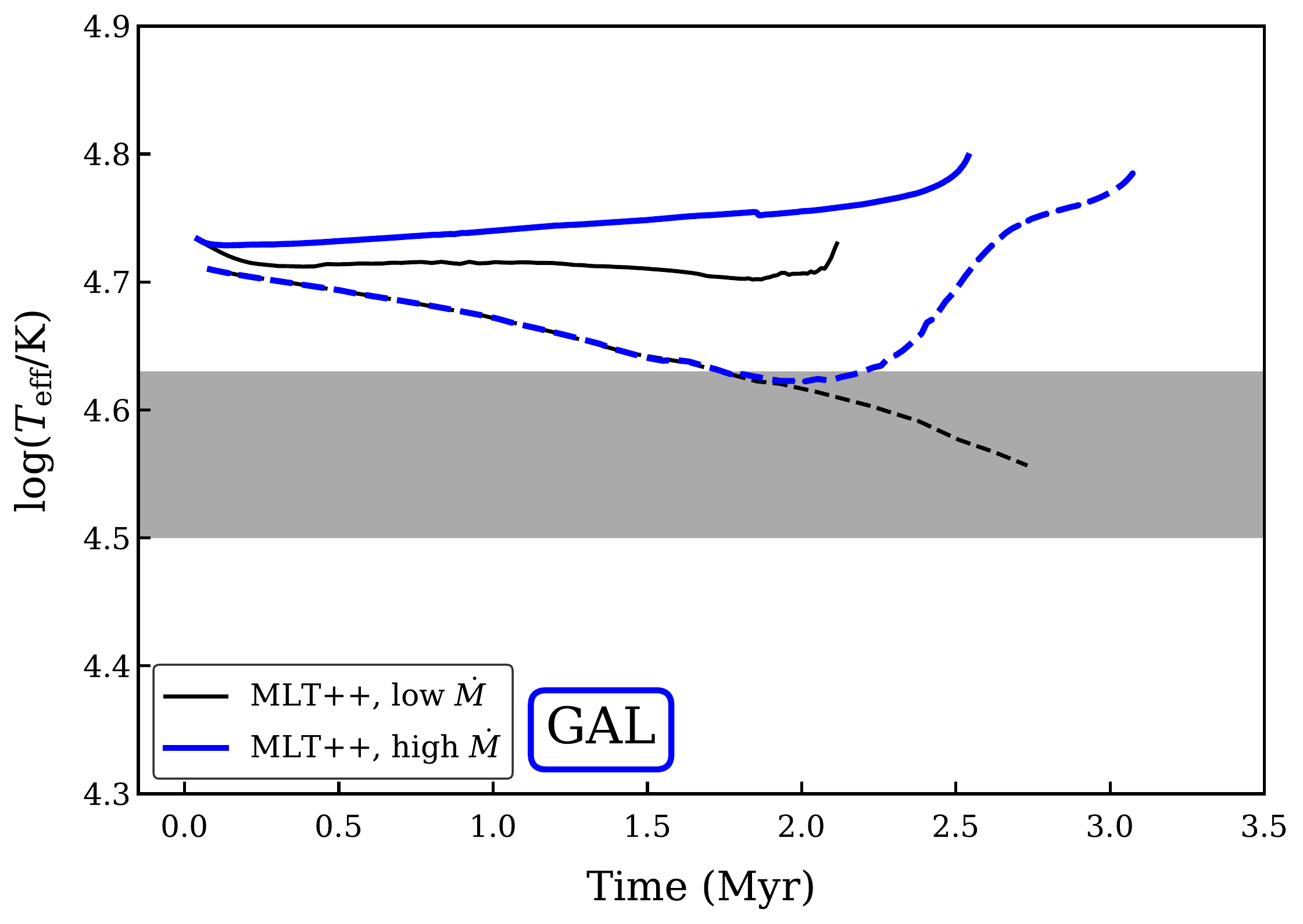}
    \caption{Temperature evolution of a 100 (dashed lines) and 400 $M_\odot$ (solid lines) model a Galactic metallicity with MLT++ on. The model inputs only differ in the mass loss as described in the text. The grey shaded region marks the observed range of temperature in the Arches cluster (log($T_\mathrm{eff}) \approx 4.5-4.6$}
    \label{fig:MLT++compare_temp_GAL}
\end{figure}

\begin{figure}
    \includegraphics[width = \columnwidth]{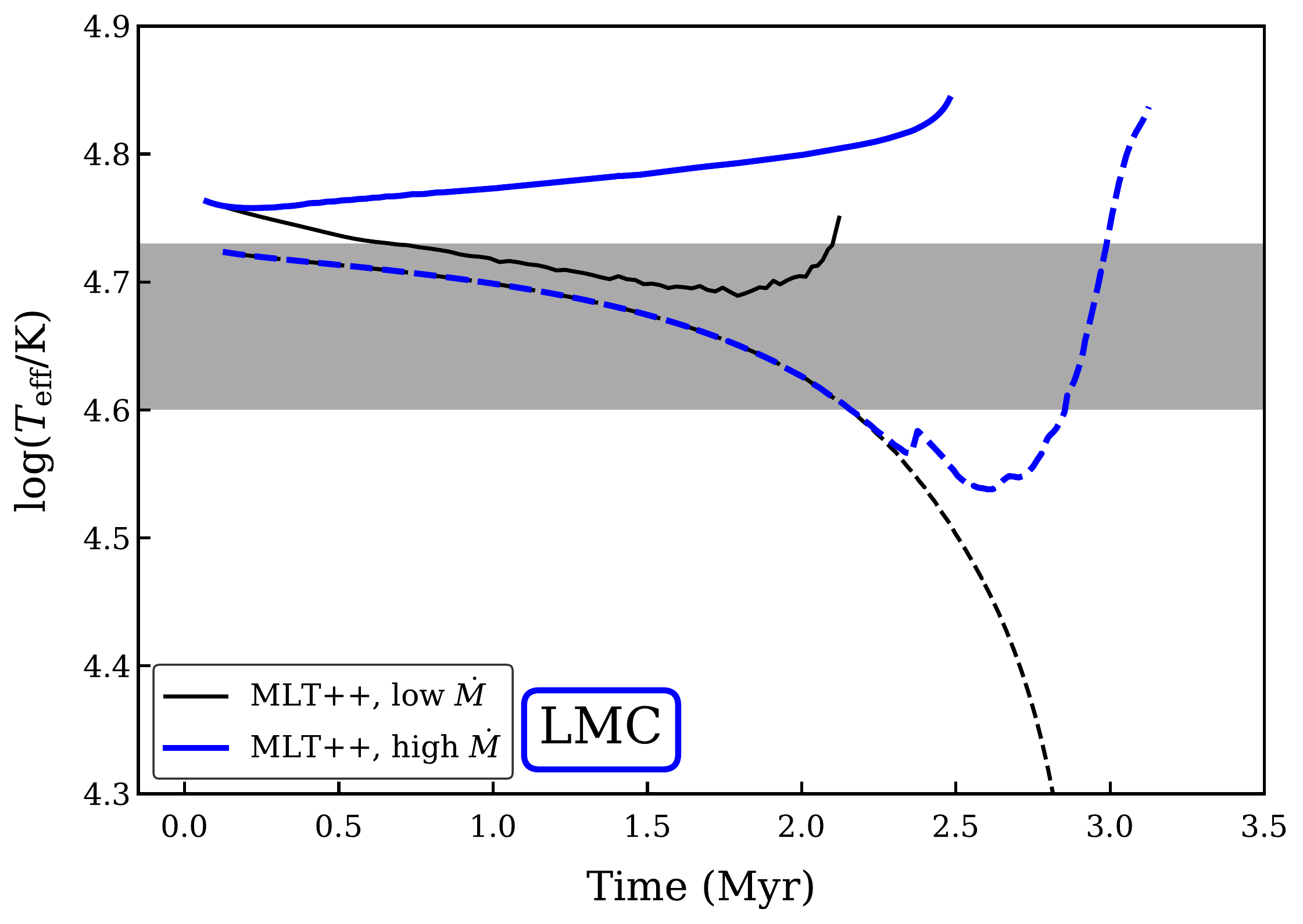}
    \caption{Same as Fig. \ref{fig:MLT++compare_temp_GAL} but at LMC metallicity. The grey shaded region marks the observed range of temperature in the 30 Dor cluster (log($T_\mathrm{eff}) \approx 4.6-4.7)$}
    \label{fig:MLT++compare_temp_LMC}
\end{figure}

\begin{figure}
    \includegraphics[width = \columnwidth]{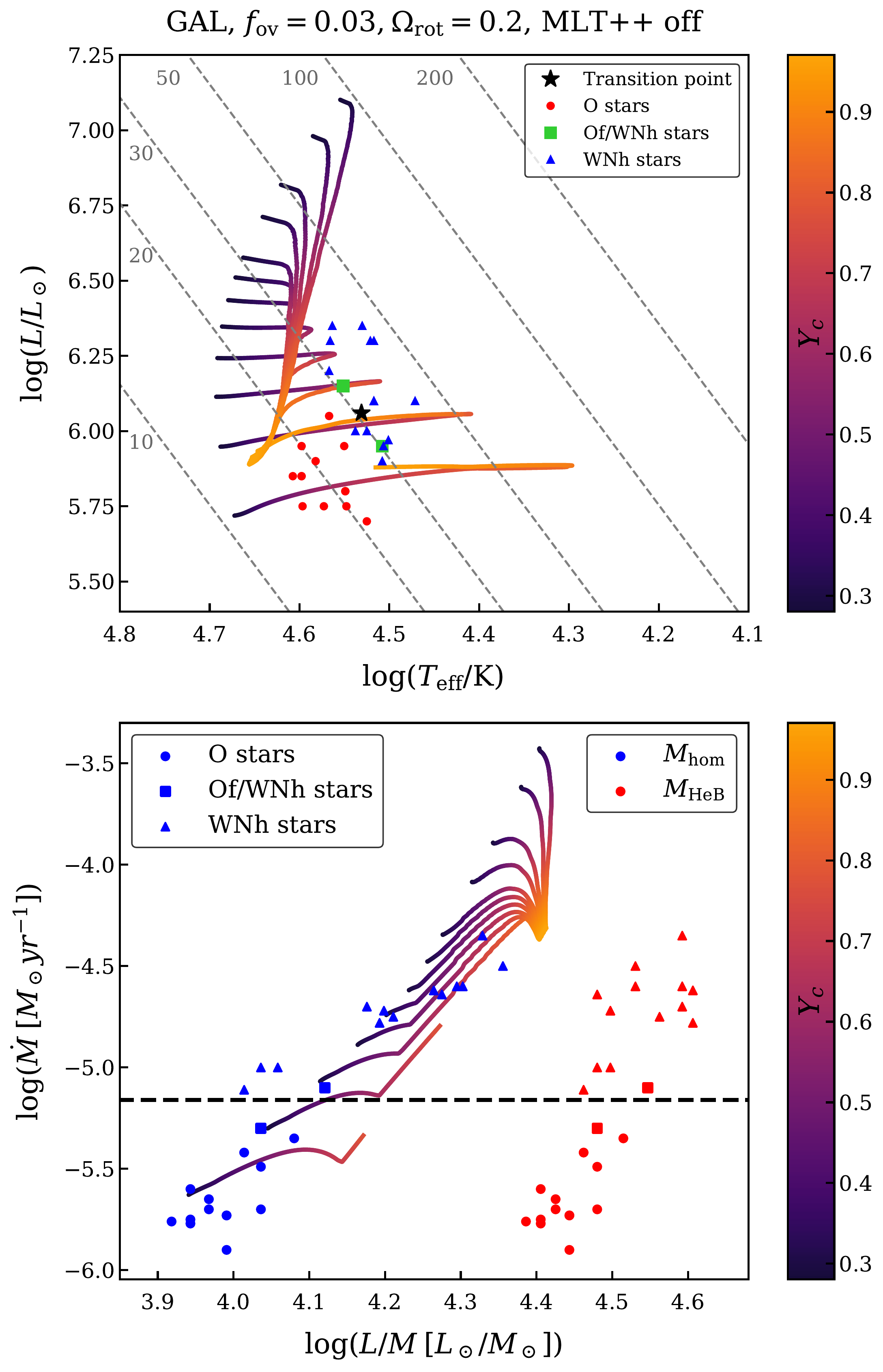}
    \caption{(Top) The stellar tracks of VMS models with initial mass 60-500 $M_\odot$  that implement the ($L/M$)-dependent mass loss. The color of the tracks correspond to the central He mass fraction. We also provide the empirical results in colored dots taken from the Arches cluster \citep{Martins2008}. (Bottom) Theoretical mass loss from our models as a function of $\mathrm{log}(L/M)$ as the star evolves. The black dashed line indicates the empirical transition mass loss derived in the Arches cluster. The stellar mass is derived from chemically homogeneous models assuming core H burning (blue) and core He burning (blue) as described in \citet{Graf2011}.}
    \label{fig:HRD_gal}
\end{figure}

We plot the temperature as a function of time in Fig. \ref{fig:MLT++compare_temp_GAL} and \ref{fig:MLT++compare_temp_LMC} at different initial metallicities corresponding to the Galaxy and the LMC. The grey shaded region marks the observed temperature range of the VMS in the two clusters. 

Lets consider the behavior of the 100 $M_\odot$ model with MLT++ on and low-$\Gamma_\mathrm{e}$ mass loss (black dashed line). At these masses, the mixing induced by MLT++ during the MS is not strong enough to counter the inflation effect. Without enhanced mass loss, the models evolve redwards. We also see that when models switch to the enhanced high-$\Gamma_\mathrm{e}$ mass loss (as the stars are close to the transition at these masses), the stars immediately evolve bluewards. Moreover at these masses, the temperature evolution is also sensitive to other processes such as overshooting and rotation.

A different behaviour is obtained in the temperature evolution of the 400 $M_\odot$ model with MLT++ and low-$\Gamma_\mathrm{e}$ mass loss (black solid line). At the highest masses, the inflation is balanced by the mixing induced by MLT++ and the model evolves at almost constant temperature, similar to the behavior seen in the previous section where enhanced mass loss reduced the luminosity keeping the inflation in check. Here efficient energy transport in the envelope directly reduces the Eddington parameters inside the star, thus balancing the inflation. 

These calculations show that the MLT++ models can reproduce only the qualitative behavior of evolving at constant temperatures. They fail to reproduce the observed temperatures in the Arches cluster, as models with MLT++ predict higher temperatures already at the ZAMS, and might hint towards either erroneous input values such as the initial metallicity taken for the Arches cluster or MLT++ incorrectly predicting the efficiency of mixing at the highest masses. Moreover, models with both enhanced mass loss and mixing (blue solid line) begin their MS at temperatures hotter than the observed temperatures and monotonously evolve bluewards. They can easily flip towards temperatures hotter than the ZAMS as demonstrated in \citet{Graf2021}. This is the effect of both high-$\Gamma$ mass loss and MLT++ which inflation is unable to counter. 

To summarize, the MLT++ routine qualitatively mimics the temperature evolution of enhanced high-$\Gamma_\mathrm{e}$ mass loss models. This leads to a degeneracy between models with increased mass loss and models with MLT++ on as they both result in qualitatively similar behavior in the temperature evolution of VMS. However the absolute temperatures predicted by MLT++ are too hot. This mismatch could be seen as an argument against MLT++.

\subsection{VMS model grid}
\label{sec:model_grid}

We now present a full grid of VMS models with varying initial mass that use our $L/M$-dependent mass loss. Fig. \ref{fig:HRD_gal} (top) shows the MS stellar tracks of stars with initial mass ranging from $60 - 500 \;M_\odot$. We over-plot the observed HR diagram locations of O (red circles), Of/WNh (green squares) and WNh stars (blue triangles) in the Arches cluster from \citet{Martins2008}. The black star symbol marks the `average' luminosity and the effective temperature of the transition stars obtained by averaging the surface properties of the Of/WNh stars (green squares). This physical transition point is used to calibrate our high-$\Gamma_\mathrm{e}$ recipe as described in Sect \ref{sec:full_mdot}. 

The mass loss as a function of $\mathrm{log}(L/M)$ for different initial masses is shown in Fig. \ref{fig:HRD_gal} (bottom). The mass loss smoothly switches - from the low to the high-$\Gamma_\mathrm{e}$ case - at the transition mass loss $\mathrm{log}(\dot{M}_{\mathrm{trans}})  = -5.16$ (black dashed line). The $\dot{M}-(L/M)$ slope also increases from $\approx 2$ to $\approx 5$ as described in Eqn. \ref{eq:optically_thin} and \ref{eq:optically_thick_LM}.

What's interesting is the change in the qualitative evolution of mass loss as a function of time with increasing initial mass. At the lowest masses ($\lesssim$ 100 $M_\odot$) presented here, the mass loss increases as the star evolves. This is due to the steady increase in the luminosity during the MS where $\mu_{\mathrm{core}}$ increases as hydrogen fuses to produce heavier He nuclei. As the star continuously undergoes mass loss, this could potentially decrease the luminosity with decreasing total mass. However the mass loss at these masses is not strong enough to yield a net reduction in the luminosity. Thus mass loss continuously increases as the star evolves. 

At the upper end of the initial mass spectrum ($\gtrsim$ 200 $M_\odot$) considered here, the VMS begin their H burning with luminosities in excess of log$(L/L_\odot) = 6.5$. The mass loss is strong enough to quickly evaporate a large fraction of the initial mass of the star, resulting in an overall decrease in the luminosity. Although one would expect the drop in the mass loss to halt the decreasing luminosity, we see a continuous drop in the luminosity as the model evolves. This is due to the high value of the absolute mass loss in the VMS mass range and the not-so steep decline in the mass loss as it is a function of the ratio $L/M$.
Both the luminosity and the total mass of the star decreases while the ratio $L/M$ remains almost a constant. The highest mass models thus see a vertical drop in the mass loss as a function of the ratio $L/M$.

Fig. \ref{fig:HRD_gal} (bottom) also plots empirical mass loss results obtained with $D=10$ in the Arches cluster as a function of $L/M$. Here $L$ is the observed luminosity and $M$ is derived by assuming chemical homogeneity. The masses for the blue symbols are derived for a given luminosity $L$ and surface hydrogen mass fraction $X_\mathrm{s}$ by assuming a fully homogeneous H burning star with $X(r) = X_\mathrm{s}$ \citep[Eq. 11,][]{Graf2011}. The red symbols are obtained by assuming a fully homogeneous helium star with $X(r) = 0$ \citep[Eq. 13,][]{Graf2011}. For a star with given $L$ and $X_\mathrm{s}$, the blue symbol represents the maximum mass or the minimum $L/M$, while the red symbol represents the minimum mass or the maximum $L/M$. 

The mass loss used above the transition point in our models assumes no explicit dependence on temperature or surface H abundance. Despite the aforementioned simplification in our modelling, there is a good agreement between the solid lines implementing our mass loss and the empirical results assuming chemical homogeneous hydrogen stars (blue symbols). As we see in \ref{sec:chem_hom}, the VMS evolve almost fully homogeneous agreeing with the results we find here. The red symbols are located farther to the right suggesting the VMS in the Arches cluster are not core He burning objects.

\begin{figure}
    \includegraphics[width = \columnwidth]{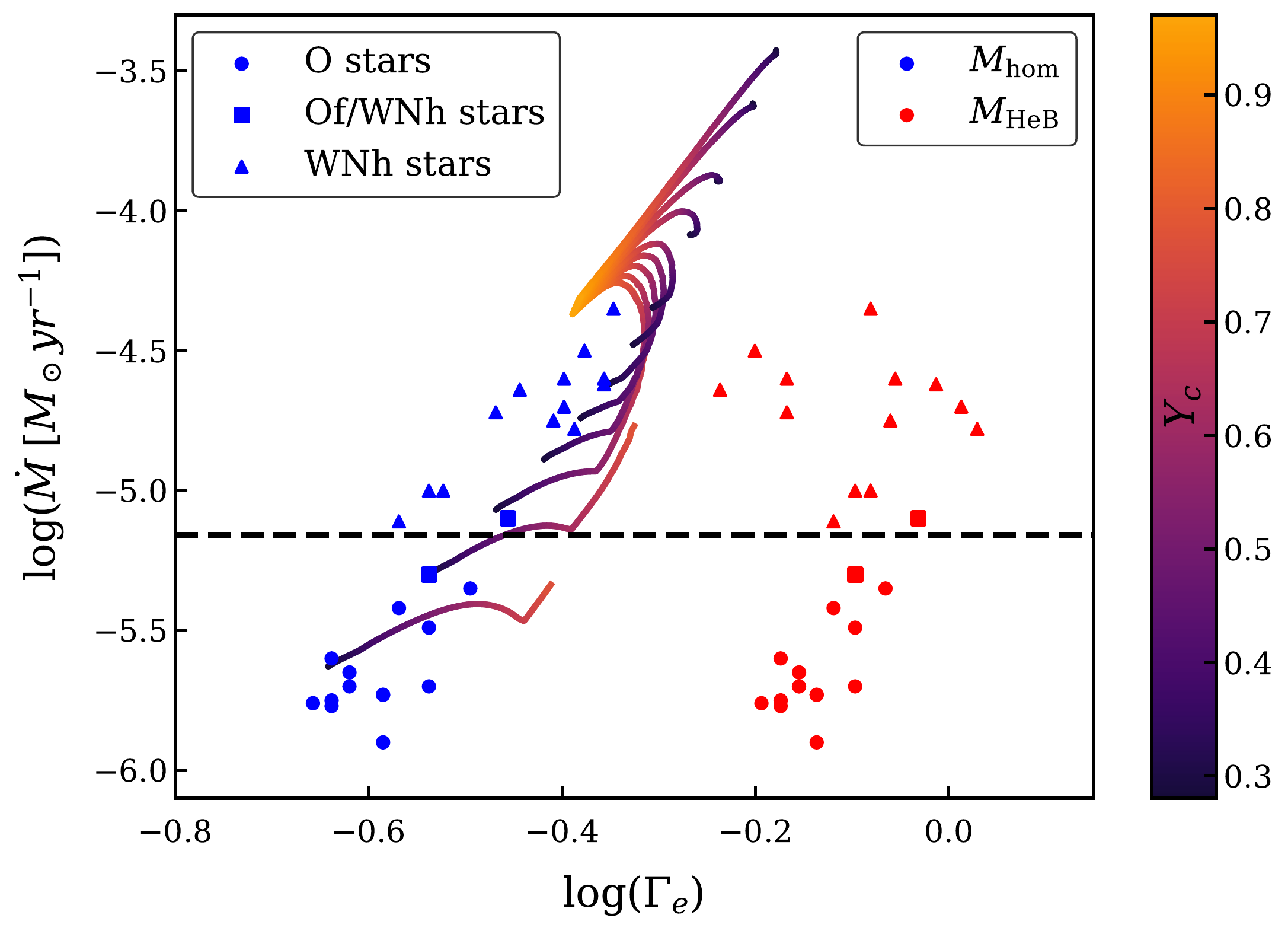}
    \caption{Mass loss used in our stellar tracks as a function of the electron scattering Eddington parameter $\mathrm{log}(\Gamma_\mathrm{e})$. The markings have the same meaning as Fig. \ref{fig:HRD_gal}.}
    \label{fig:mdot_gamma_gal}
\end{figure}

In Fig. \ref{fig:mdot_gamma_gal}, we plot the mass loss as a function of the electron scattering Eddington parameter $\Gamma_\mathrm{e} \propto (1+X_\mathrm{s}) (L/M)$. The dots and black solid tracks have the same meaning as the previous figure. Once again the blue symbols have the minimum $\Gamma_\mathrm{e}$ for a given value of $L$ and $X_\mathrm{s}$, while the red symbols have the maximum $\Gamma_\mathrm{e}$ assuming  $X_\mathrm{s} = 0$. The stars below the transition evolve similarly to the previous figure as $X_\mathrm{s}$ remains close to 0.7 throughout the entire evolution and $(L/M) \propto \Gamma_\mathrm{e}$. However, the surface abundances closely follow the central abundances above the transition point. This is due to chemically homogeneous evolution, which we discuss in the following section. Combined with the steep drop in luminosity, $\Gamma_\mathrm{e}$ drastically drops as the star evolves. A potential consequence of such a decrease in $\Gamma_\mathrm{e}$ on the radius evolution will be discussed in Sec. \ref{sec:discussion}.

\subsection{Chemically homogeneous evolution}
\label{sec:chem_hom}

Another interesting feature evident from our models are the overlapping tracks of models beginning with high enough initial masses, $M_{\text{init}} \gtrsim 200 \; M_\odot $ --  specifically for the later half of their MS. This behavior can be understood by carefully studying the internal profile of these stars and their total mass evolution.

\begin{figure}
    \includegraphics[width = \columnwidth]{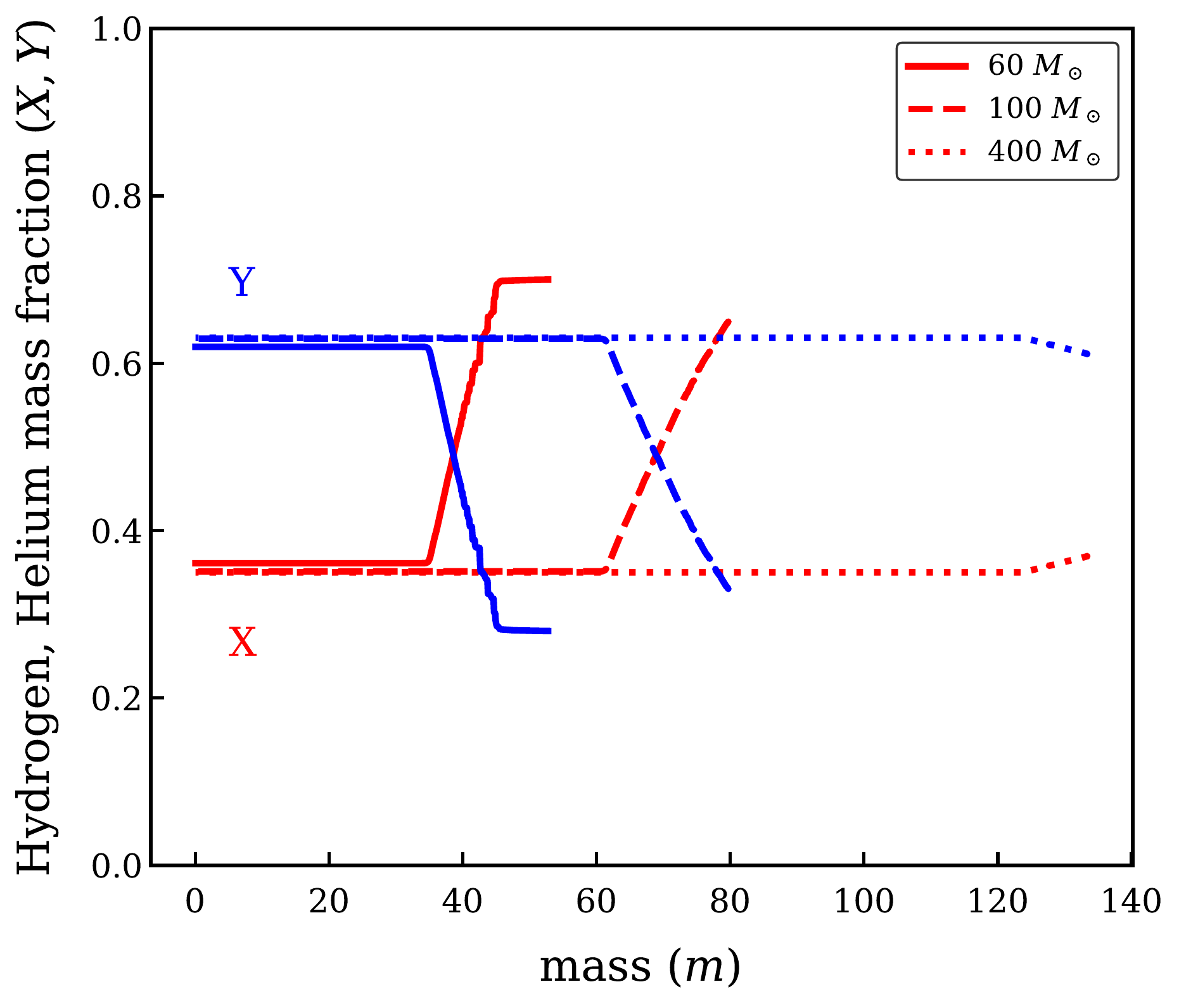}
    \caption{Internal profiles of the hydrogen (X, red) and helium (Y, blue) mass fractions of three models - 60, 100, 400 $M_\odot$. The profiles are taken when the $X_\mathrm{c}$ equals 0.35, this is half way through the main sequence.}
    \label{fig:internal_profile}
\end{figure}

The internal abundance profiles for three selected initial mass models - $60, 100$ and $400 \;M_\odot$ are shown in Fig. \ref{fig:internal_profile}. These models are representative of stars below the transition, near the transition and above the transition and show a gradual change in the ratio of the convective core mass to the total mass (we obtain the convective core mass fraction as a function of the initial mass in Appendix \ref{appendix:B}). The profiles are taken at halfway through the MS when the $X_\mathrm{c}$ equals $0.35$. 

The horizontal portions of the profile denote the well mixed, convective regions inside the star. The 400 $M_\odot$ model has a very large convective core mass to total mass ratio ($\approx 1$). The growing importance of radiation pressure with increasing initial mass (see below) which lowers the adiabatic temperature gradient and the increasing luminosity (technically the $l/m$ ratio) that needs to be transported outwards to support the gravity are both responsible for the increase in the convective core mass fraction as a function of the initial mass. Combined with the high mass loss from the optically-thick wind of VMS, the internal chemical profile of these stars are fully homogeneous.

VMS with $M_{\text{init}} \gtrsim 200 \; M_\odot$ undergo chemically homogeneous evolution throughout their main sequence. We briefly discuss the analytical description to obtain the convective core to total mass ratio of a star of given initial mass $M_{\text{init}}$. The full mathematical expressions of the quantities discussed here and a direct comparison with values obtained from detailed stellar evolution models is in Appendix \ref{appendix:B} (see Fig. \ref{fig:beta_model_theory} and \ref{fig:zams_core_size}). 

The growing importance of radiation pressure in massive stars is captured by the quantity $\beta (r)$ defined by the ratio of gas pressure to the total pressure at a location $m(r)$ inside the star. We closely follow the homology arguments of \citet{Maeder2012} in Appendix \ref{appendix:B} (see Eq. \ref{eq:beta_mass_relation}) and $\beta$ has the following scaling with the mass
\begin{equation}
\begin{split}
\beta \sim \dfrac{1}{\sqrt{M}}
\end{split}
\label{beta_mass}
\end{equation}
In other words, the quantity ($1-\beta$) proportional to the radiation pressure increases with the initial mass of the star.

The convective core size of a star is determined at the location $m(r)$ where the temperature gradient required for radiation diffusion to transport the stellar luminosity outwards, $\nabla_{\mathrm{rad}} \sim \chi l/m$ equals the adiabatic temperature gradient $\nabla_{\mathrm{ad}}$. For regions where $\nabla_{\mathrm{rad}} > \nabla_{\mathrm{ad}}$ like in the central region of a massive star, the entire luminosity cannot be transported only by radiation and convection sets in. Whereas for $\nabla_{\mathrm{rad}} < \nabla_{\mathrm{ad}}$, radiation alone can transport the entire energy produced. 

Solving then for $m(r, \nabla_{\mathrm{rad}} = \nabla_{\mathrm{ad}})$ (see Eq. \ref{eq:m_core_size}) gives the size of the convective core. The convective core mass fraction increases with initial mass, reaching values greater than $\approx 0.9$ at the ZAMS for $M_{\text{init}} \gtrsim 200 \; M_\odot$. The models with $M_{\text{init}} \gtrsim 200 \; M_\odot$ remain fully mixed throughout the MS regardless of their rotational and overshooting inputs.

A consequence of having a fully mixed star is the insensitivity of the evolution of VMS to processes that affect the core size such as overshooting and rotation. Above $M_{\text{init}} \approx 200 \; M_\odot$, the effects of varying overshooting and rotation on their evolution is negligible and the evolution is completely dominated by mass loss. Thus the arguments made here for one set of overshooting and rotation parameters hold true for other values of $f_\text{ov}$ and $\Omega/\Omega_{\text{crit}}$ within a reasonable range.

In addition to the fully mixed internal profile of VMS, they also have very similar masses towards the end of their MS. This is a consequence of the steeper $\dot{M}-L$ slope. The mass loss is strongest at the highest initial masses, strong enough to quickly evaporate a large portion of the initial mass. Thus the overlapping of the evolutionary tracks in the HR diagram is due to these objects having very similar masses and chemically homogeneous profiles. This also means the theoretical mass loss used in our models with overlapping tracks should predict very similar values.

The VMS all have very similar masses for age $\gtrsim 1.5$ Myr, thus proving the determination of their initial masses by tracing back the evolution almost impossible \citep{Vink2018, Higgins2022}. The VMS models all reach a final mass of $\approx 30 M_\odot$ at the end of main sequence, and also have very similar surface properties. It is thus nearly impossible to know the initial conditions of VMS by knowing their current mass and their location in the HRD. 

This degeneracy in the initial mass can perhaps be broken by the observed surface hydrogen content. With the well mixed convective core encompassing almost the entire star, one would expect the surface abundances to closely follow the central values. Thus for $M_{\text{init}} \gtrsim 200 \; M_\odot$, the observed surface abundances behaves like a `clock' and can be a proxy for the age of the star. We use this concept to explore the possibility of determining the initial mass of the star in \citet{Higgins2022}.

\begin{figure}
    \includegraphics[width = \columnwidth]{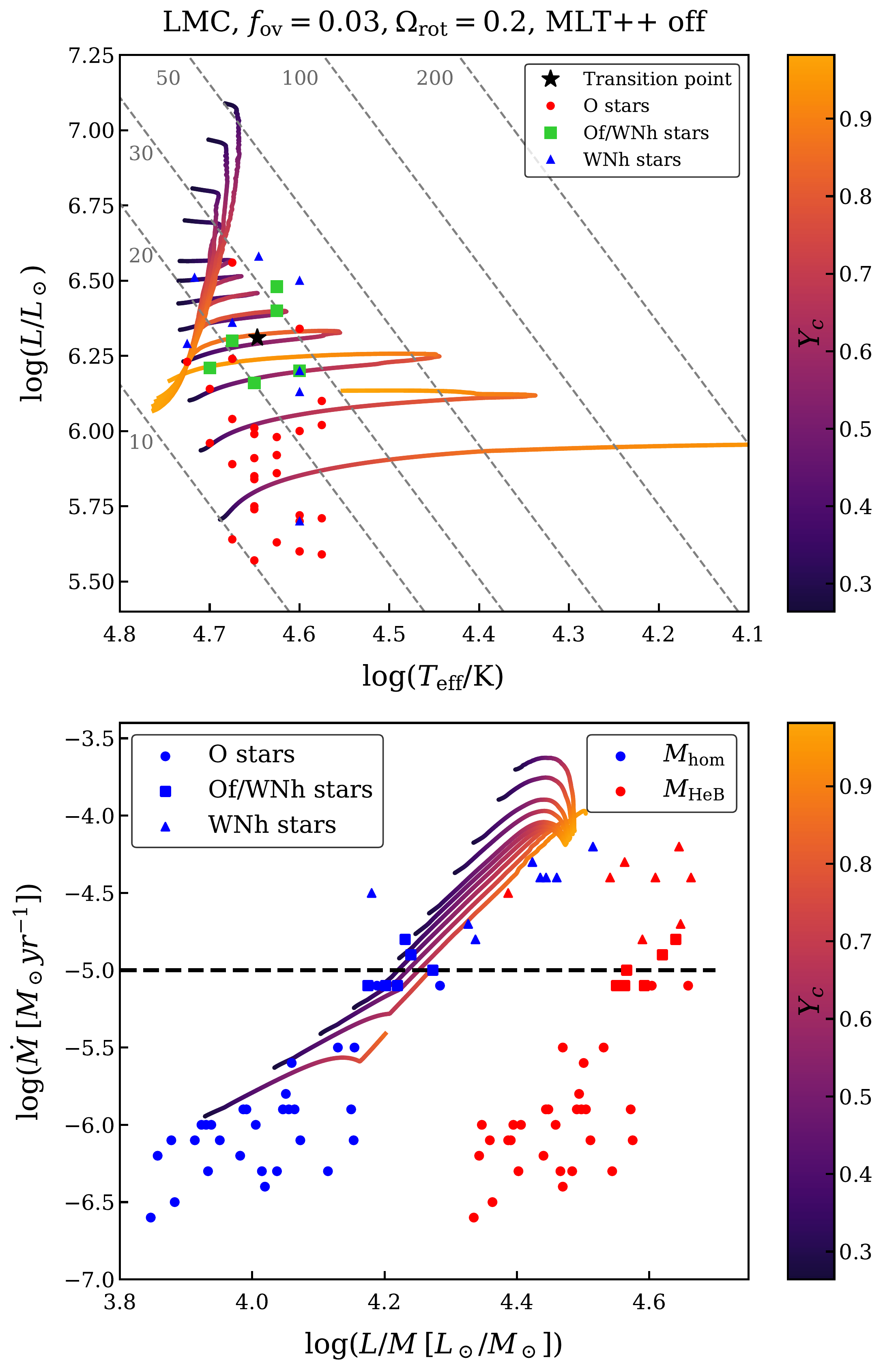}
    \caption{Stellar tracks of VMS models at LMC metallicity. The empirical results are taken from \citet{Best2014}. The black star symbol is now an average of six Of/WNh transition stars marked as green squares. The transition mass loss (bottom) is marked by the black dashed line. All other inputs are identical to Fig.\ref{fig:HRD_gal}.}
    \label{fig:HRD_lmc}
\end{figure}

\subsection{Effect of metallicity}
\label{sec:metallicity_effect}

In this subsection, we study the effects of initial metallicity on the evolution of VMS. As discussed in Sect. \ref{sec:tmr}, we have empirical results of VMS in the 30 Dor cluster in the LMC, and a transition feature at a different host metallicity. We directly compare the Galactic models to the models at LMC metallicity ($\sim$  half Galactic) computed here. Fig. \ref{fig:HRD_lmc} (top) shows the HR diagram of the LMC models, again with overshooting of $f_\text{ov} = 0.03$ and rotation of $\Omega/\Omega_{\text{crit}} = 0.2$ allowing a direct comparison with Fig. \ref{fig:HRD_gal}. The mass loss as a function of $L/M$ used in these models are plotted in Fig. \ref{fig:HRD_lmc} (bottom), along with the empirical results from \citet{Best2014}. The colored symbols in both the figures have the same meaning as Fig. \ref{fig:HRD_gal}.

The chemically homogeneous evolution for stars above a certain initial mass is also seen at reduced metallicity. The temperatures of these lower-$Z$ stars are slightly hotter ($\approx 0.1$ dex). This is the effect of both the reduced opacities near the surface and the lowered CNO abundances in the core \citep{Farell2021}. However, the predicted mass loss is inherently lower due to the lower Fe content  -- with exceptions we discuss below --  that is responsible for setting the mass loss in the inner wind region below the sonic point. Thus models near the transition evolve towards slightly cooler temperatures compared to the Galactic models. Only then the high-$\Gamma_\mathrm{e}$ mass loss becomes strong enough to strip the envelope of the star and cause them to evolve bluewards in the HRD.

Once again, we get a degeneracy in the initial mass after an age $\approx 1.5$ Myr. The decrease in the total mass in the beginning is steeper at Galactic metallicity due to higher metallicity. However, the drop in luminosity is larger at higher metallicity. The Galactic models towards the end of their main sequence evolve at lower luminosities compared to the LMC models. The effect of $L$ on the mass loss is stronger than the effect of $Z$, resulting in a higher predicted mass loss at lower metallicity towards the end of main sequence. Thus the decrease in the total mass is steeper in the beginning and shallower towards the end of MS at high $Z$ and vice versa at lower $Z$. These two opposing effects result in the TAMS masses for the LMC models ($\approx 40 \;M_\odot$) being similar to the Galactic ones ($\approx 30 \;M_\odot$). A degeneracy exists in initial metallicity as well.

\section{Discussion}
\label{sec:discussion}

In Sec. \ref{sec:temp_evolution_VMS} we presented VMS test models with input mass loss that implicitly depended on $X_\mathrm{s}$ through a steep dependence on $\Gamma_\mathrm{e}$ and compared that to a model with a pure $L/M$-dependent mass loss with no $X_\mathrm{s}$ dependence. Although the absolute mass loss predicted by the two different dependencies differ only by an order of 0.1 dex, the two models had noticeable qualitative differences in the HR diagram. Here we discuss the differences in the evolution of the two models and explain them by carefully examining their internal radius profiles.

\begin{figure}
    \includegraphics[width = \columnwidth]{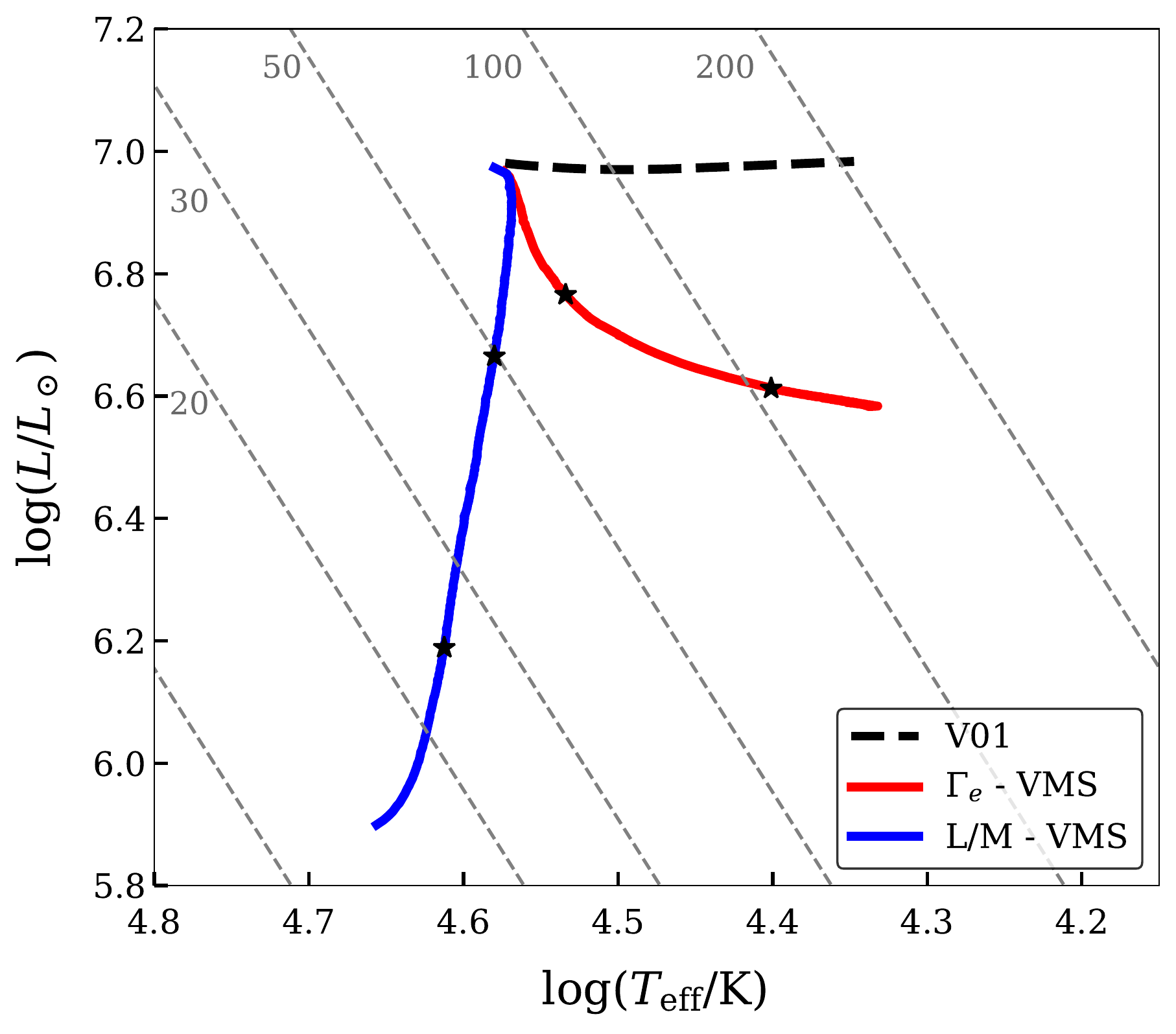}
    \caption{Comparison of 400 $M_\odot$ VMS main sequence evolution using different mass-loss dependencies on the stellar properties (see Sec. \ref{sec:temp_evolution_VMS} for details of individual test model). Black star symbols mark the ages 1 and 2 Myr along the stellar track. All the models use an input overshooting efficiency of $f_\text{ov} = 0.03$ and rotate at 0.2 times the critical velocity at ZAMS. The absolute mass loss used is plotted in Fig.  \ref{fig:mdot_COMPARE}.}
    \label{fig:HRD_COMPARE}
\end{figure}

\begin{figure}
    \includegraphics[width = \columnwidth]{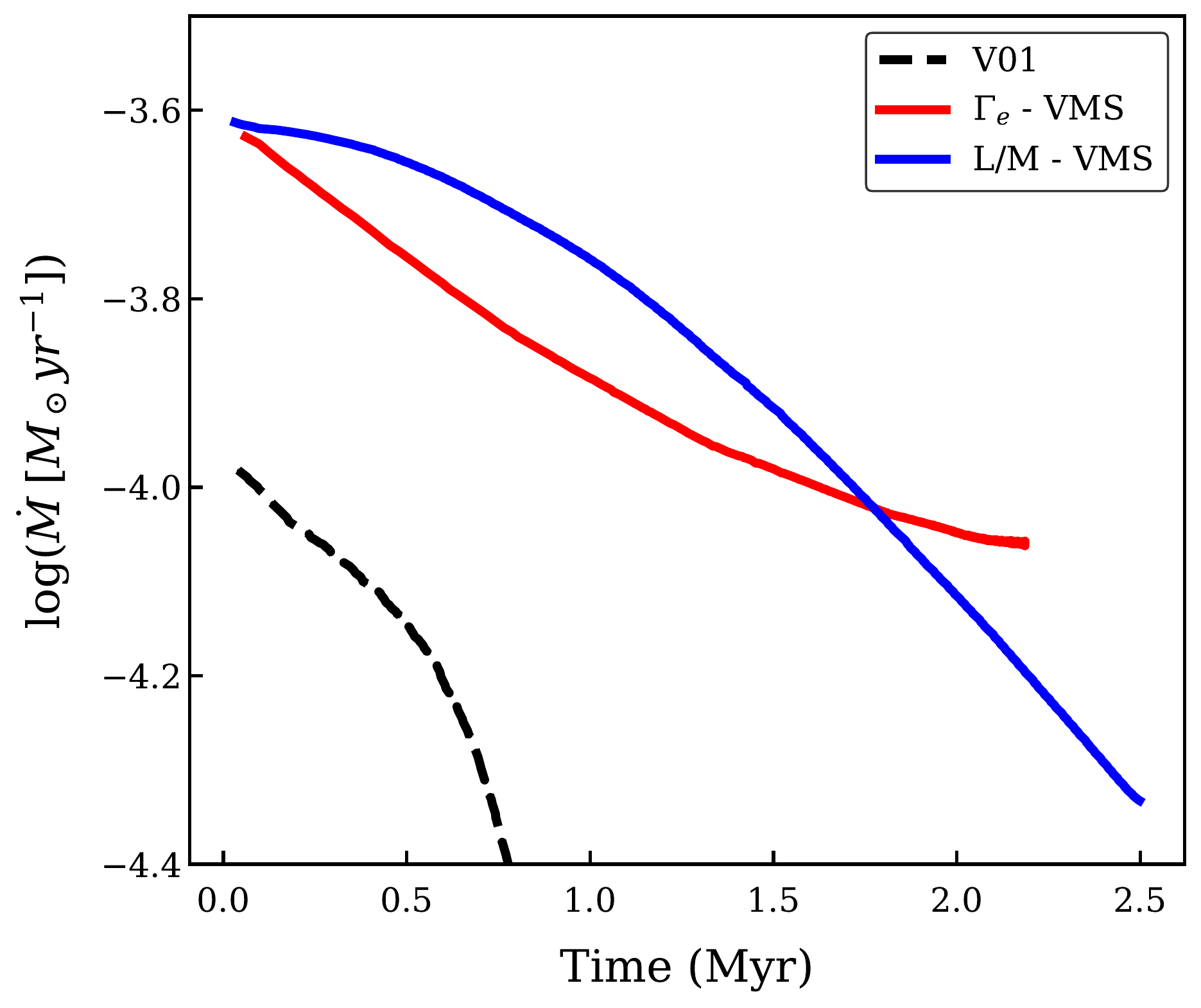}
    \caption{Absolute mass loss as a function of time used in our test models. The black dashed line follows the low-$\Gamma_\mathrm{e}$ mass loss from Eq. \ref{eq:optically_thin}, and the red and blue solid lines follow the mass loss predicted by Eq. \ref{eq:optically_thick_gamma} and \ref{eq:optically_thick_LM} respectively.}
    \label{fig:mdot_COMPARE}
\end{figure}

In Fig. \ref{fig:HRD_COMPARE} and \ref{fig:mdot_COMPARE}, we schematically compare the HR diagram evolution of Galactic models and the corresponding absolute mass loss used for the three different cases - V01 (black), VMS-$\Gamma_\mathrm{e}$ (red) and VMS-$L/M$ (blue) (see Sec. \ref{sec:full_mdot} and \ref{sec:temp_evolution_VMS} for the details of these models). This color scheme is adopted throughout this section.

The V01 model evolves at almost constant luminosity and inflates horizontal towards the red part of the HRD. Hence it cannot explain the narrow temperature range of VMS. Such a redward expansion is typical of classical main-sequence O stars. With V01 rates, a 400 $M_\odot$ VMS loses mass through an optically-thin wind and evolves qualitatively similar to a canonical O star. Both theory and empirical evidence have shown that the absolute mass loss predicted by \citet{Vink2001} falls short for stars above the transition.

\begin{figure}
    \includegraphics[width = \columnwidth]{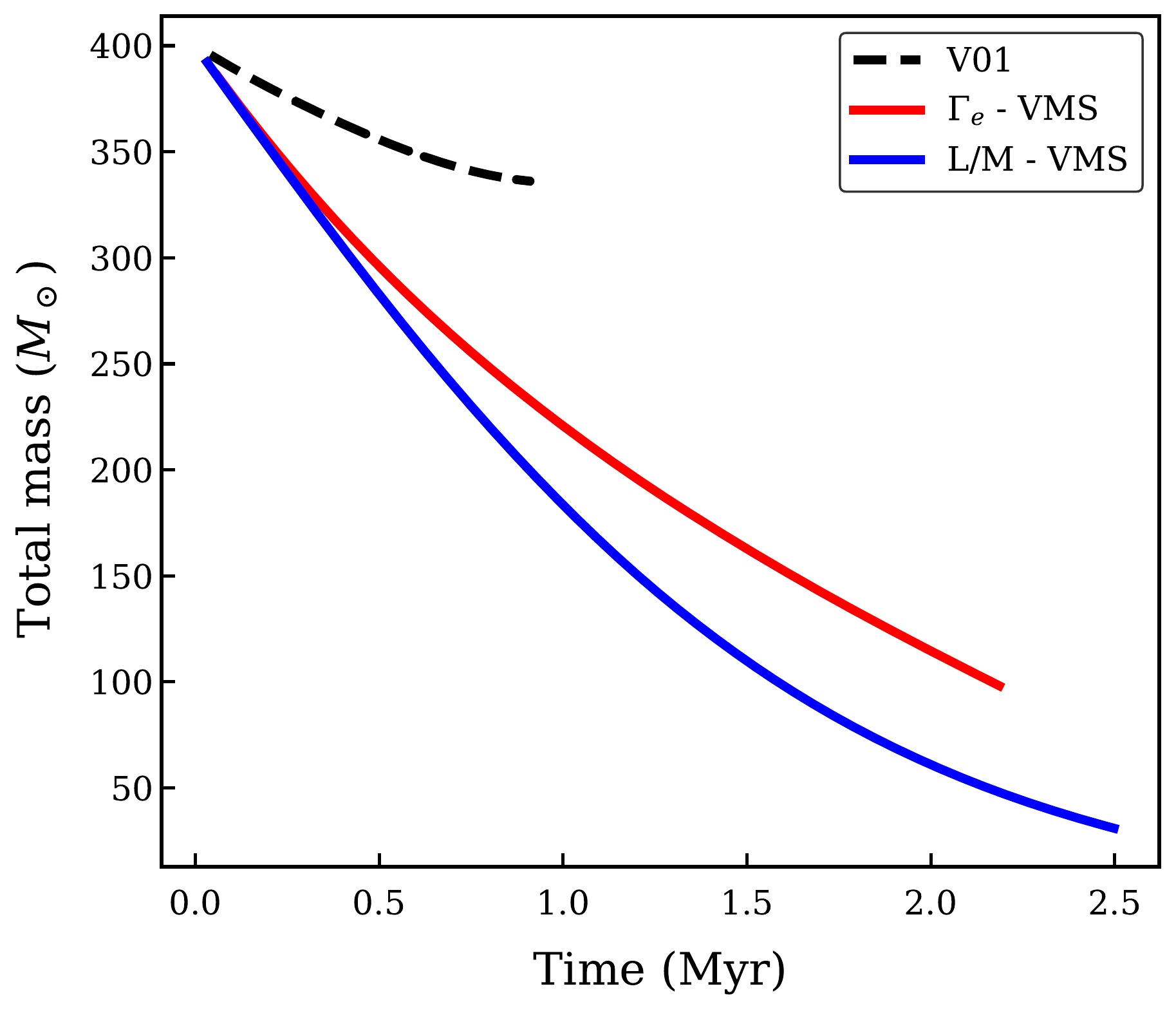}
    \caption{The total mass evolution of a 400 $M_\odot$ as a function of time. Notice the much steeper decline in the mass when employing an optically-thick Wolf-Rayet type wind that scales steeply with the luminosity.}
    \label{fig:mass_COMPARE}
\end{figure}

The effect of the under-prediction in the mass loss by \citet{Vink2001} above the transition is also reflected in Fig. \ref{fig:mass_COMPARE} where we plot the total mass evolution as a function of time. The gradient for the V01 model (black) is significantly shallower when compared to models  that use a steeper $\dot{M}-L$ relation in the high-$\Gamma_\mathrm{e}$ regime (red and blue). 

The red and blue lines both implement a steeper $\dot{M}-L$ slope only differing by their dependence on the $X_\mathrm{s}$. An immediate qualitative difference we notice between these models and the V01 model is a drop in stellar luminosity during the evolution. Despite the increasing $\mu_\mathrm{core}$, the red and blue models see a drop in luminosity, more so in the blue model (see below). This drop in luminosity is caused by the higher absolute mass loss used in these models, as strong enough mass loss can quickly strip a large fraction of the initial stellar mass resulting in a net drop in luminosity.

The MS lifetime can also be affected by the changes in the luminosity evolution of VMS. The MS nuclear burning timescale can be approximately given as 
\begin{equation}
\begin{split}
t_\mathrm{MS} = \dfrac{E_\mathrm{nuc}}{L} = \dfrac{f_\mathrm{H} (f_\mathrm{m} M) c^2}{L}
\end{split}
\label{eq:MS_lifetime}
\end{equation}
where $f_\mathrm{H}$ is the fraction of rest mass that is converted to energy during hydrogen fusion ($f_\mathrm{H}$ = 0.007) and $f_\mathrm{m}$ is the fraction of the total mass of the star that is available as fuel. Given a mass-luminosity relation $L \sim M^{x}$ where $x>1$, the highest mass stars despite having a larger fuel supply burn through it faster due to their higher luminosity. For the two models considered here, there is a significant difference in the final luminosities during their evolution, sufficient enough to result in a half a million year difference in their ages, of the order of 20$\%$ of the total MS lifetime.

Another striking difference is seen in the HR diagram evolution of the red and blue curves. These models differ only in the mass-loss rate owing to the different treatment of their hydrogen dependence. As the VMS evolves the surface hydrogen content closely follows the central value, as discussed in Sec. \ref{sec:chem_hom}. As the surface H enters into the mass loss through the term $\Gamma_\mathrm{e}$, the model with $\Gamma_\mathrm{e}$-dependent VMS wind will have its overall mass loss suppressed by the decreasing surface H as seen in Fig. \ref{fig:mdot_COMPARE}. However the difference in the mass loss is $\approx$ 0.1 dex, which is just about the order of uncertainty in the mass loss typically derived for massive stars. 

The 0.1 dex difference in the mass loss can lead to qualitatively different evolution of VMS, specifically the redward or blueward evolution in the HRD. Massive stars typically evolve redward during the MS, where the outer envelope layers must expand to lower temperatures as the convective core gradually contracts in radius so as to keep the star in thermal and hydrostatic equilibrium. However the blue VMS model considered here has a qualitatively different behavior and evolves at almost constant temperatures, but drastically reducing in radius. Thus both the luminosity and effective temperature of VMS are highly sensitive to the input mass loss. Below we discuss the possible reasons for such a stark difference in their evolution.

The absolute mass loss of VMS is atleast an order of magnitude higher than stars below the transition. Even small uncertainties at these masses can lead to significantly different final luminosities as seen previously in Fig. \ref{fig:HRD_COMPARE}. The two models we test here quantitatively differ in their final luminosity by $\sim$ 0.6 dex, a factor of four. The total mass evolution of the red and blue models are noticeably different. Although the difference at the end of MS is of the order of $10\%$ of the initial mass, the final mass of the red model is almost double the blue model. This difference in their final masses is  caused by a difference of $\approx 0.1 $ dex in the mass loss cumulated throughout the evolution. This difference in the luminosity drop can have consequences for the radius evolution of VMS as discussed below.

\begin{figure}
    \includegraphics[width = \columnwidth]{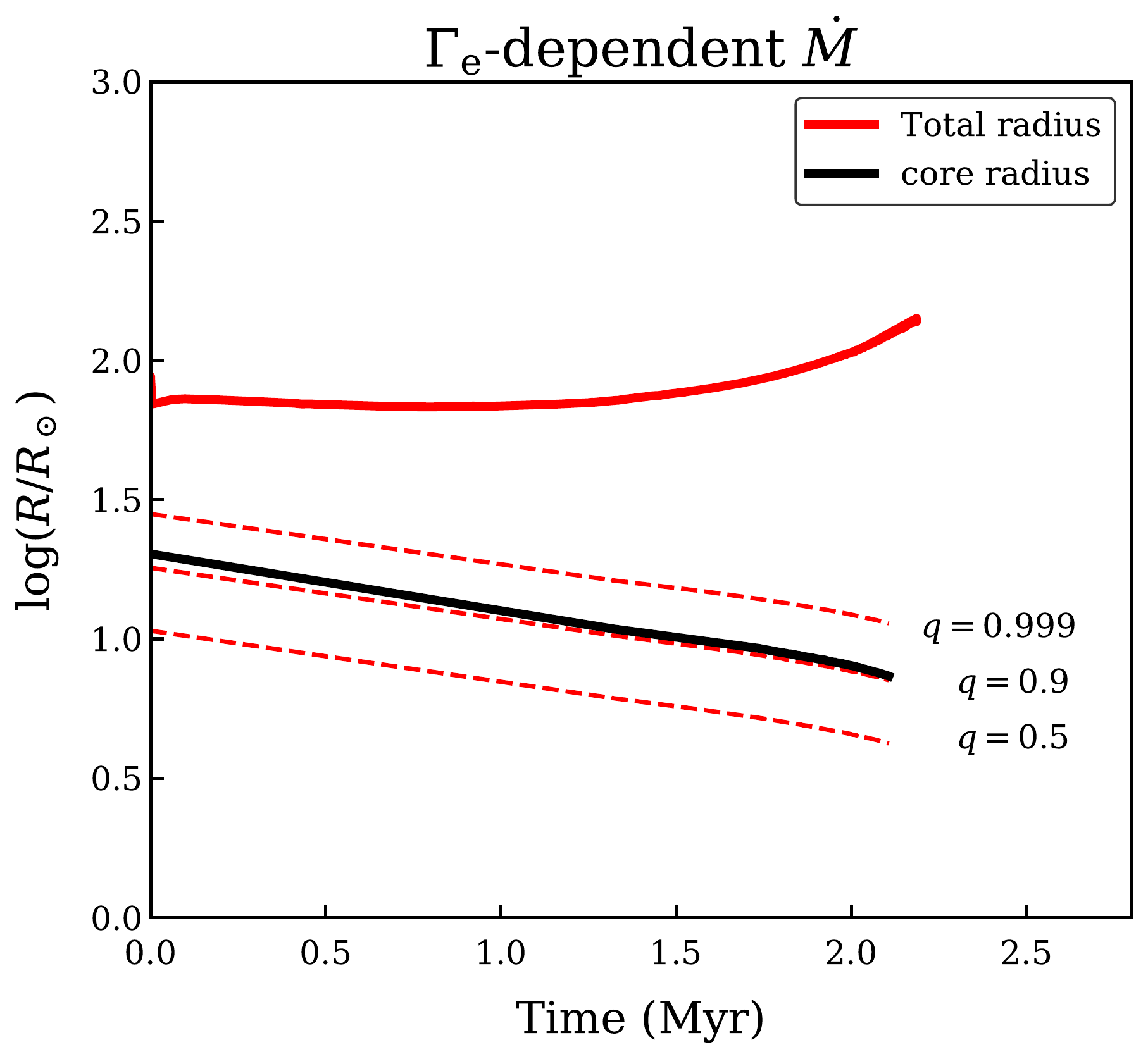}
    \caption{Total (red solid) and core radius (black solid) evolution of a 400 $M_\odot$ model with $\Gamma_\mathrm{e}$-dependent mass loss. Radius of constant $q(r) = 0.5, 0.9, 0.999$ lines are also plotted in dashed red.}
    \label{fig:radius_gamma}
\end{figure}

\begin{figure}
    \includegraphics[width = \columnwidth]{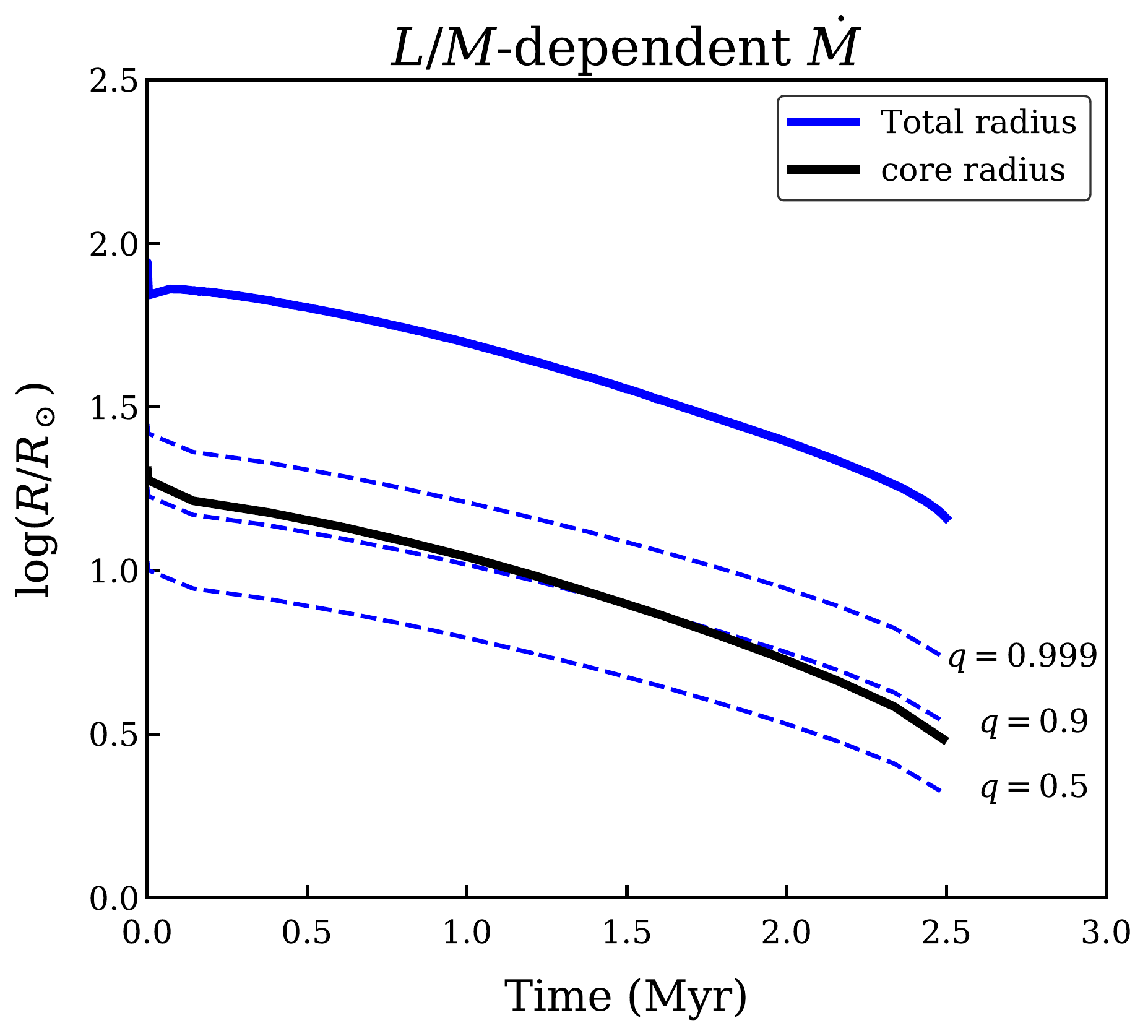}
    \caption{Same as Fig. \ref{fig:radius_gamma}, but for $L/M$-dependent mass loss VMS model.}
    \label{fig:radius_LM}
\end{figure}

To understand the radius evolution of VMS, we probe the internal radius coordinate of constant $q(r) = m(r)/M_\mathrm{tot}$ lines in the two test models to understand the radius evolution of VMS and how they differ from massive stars below the transition. Fig. \ref{fig:radius_gamma} and \ref{fig:radius_LM} show the evolution of constant $q(r) = 0.5, 0.9, 0.999$ lines along with the convective core (black) and total radius of a 400 $M_\odot$ initial mass model that only differ in their treatment of mass loss hydrogen dependence.

The convective core radius decreases in either case and follows the $q =  0.9$ line throughout the evolution. We see the convective core size recede during the MS as the electron scattering opacity in the core decreases with decreasing hydrogen. Unlike canonical O stars, the high core mass fraction of VMS can lead to the influence of the contracting core extend all the way till $q(r) = 0.999$. As seen in both the figures, the $q(r) = 0.999$ line contracts along with the core. The evolution of the inner 0.999$M_\mathrm{tot}$ and the internal temperature profile of the two models considered here are unaffected by the tiny differences in the input mass loss.

There is a qualitative difference in the evolution of their total radius. The outermost thousandth of the total mass at the ZAMS occupies the outer half of the total radius of the star. The extent of radius inflation is closely related to proximity to the Eddington limit. In Sec. \ref{sec:model_grid} we showed that for VMS using a steeper $\dot{M}-L$ dependence, the luminosity can have a net decrease during the MS causing a decrease in $\Gamma_\mathrm{e}$. A factor of four difference in the final luminosities of the red and blue models and only a factor of two difference in their final masses corresponds to a difference in their $L/M$ by a factor of two. The inflation effect is thus much stronger for the red model which requires a higher luminosity to be transported outwards, and consequently a more extended radius.

The changes in the internal profile of the two models is confined to less than the outermost thousandth of the total mass of the star. This inflated layer that contains a small fraction of the total mass is then highly sensitive to the mass loss from the surface. The lower luminosity that needs to be transported outwards in the blue model suppresses the radius inflation and combined with chemically homogeneous evolution can cause a decrease in the radius during the evolution.

\section{Summary}
\label{sec:conclusion}

In this paper we studied the main sequence evolution of Very Massive stars using the 1D stellar evolution code MESA, up to initial masses of 500 $M_\odot$. We use the physical transition mass loss from \citet{Vink2012}, a model-independent way to constrain the VMS mass loss and specifically investigate the effects of different mass-loss dependencies on the temperature evolution of VMS. The results and conclusions are summarized below:
\begin{itemize}
    \item[--] The observed temperatures of VMS in the Arches and the 30 Dor clusters are confined to a narrow range of values and increase with decreasing Z. We compare two different mass loss recipes that scale as a function of the Eddington limit or the luminosity-over-mass ratio, that use two extreme dependencies on the surface H abundance. The $\Gamma_\mathrm{e}$-dependent mass loss has a steep implicit dependence on the surface H through the number of free electrons available for scattering. On the other hand, an $L/M$-dependence does not scale with the surface H. We use this extreme case of no dependence to understand the qualitative behavior of VMS evolution for which the effect of $L/M$ on mass loss dominates over that of surface H. 
    \item[--] A constant observed $T_\mathrm{eff}$ might be interpreted as evidence for a burst of star formation in a cluster with the stars in the cluster all at the same age, however it might not be that simple for the highest masses. Both the effects of substantial envelope inflation and higher mass loss can vary with initial mass. For example, the most massive star will inflate to lower $T_\mathrm{eff}$ than the least massive star in the population, while the effects of higher mass loss can "flip" the higher mass star towards hotter temperatures. Vertical evolution where the luminosity drops steeply can naturally account for constant temperatures of VMS by self-regulating and balancing the effect of mass loss and inflation throughout the evolution and also explain the trend in the temperature as a function of host Z. While it is possible to explain the constant temperatures using models that evolve horizontally, it becomes quite challenging to balance these two effects - over the entire mass and metallicity range.
    \item[--] At the highest masses, mass loss completely dominates the evolution of VMS. Even small uncertainties of order 0.1 dex in the input mass loss can qualitatively change the luminosities and temperatures at the end of H burning, consequently affecting later stages. We find that these changes are limited to the outer thousandth of the stellar mass, where even small changes in the mass can affect the surface properties drastically. The $L/M$-dependent mass loss models explored in this paper see a considerable drop in the Eddington parameter thus suppressing the effects of radius inflation. This causes the radius to decrease during the MS, a qualitatively different behavior as opposed to typical VMS models that inflate during the MS.
    \item[--] We produce a grid of VMS models at Galactic and LMC metallicity with our new $L/M$-dependent mass loss. The VMS undergo chemically homogeneous mixing above $M_\mathrm{init} \gtrsim 200 M_\odot$, and evolve with very similar temperatures and luminosities. This creates a degeneracy where given the current location in the HR diagram and the current mass of the star, it is near impossible to determine its initial mass. This degeneracy can potentially be broken if the current surface hydrogen is well constrained \citep[see][]{Higgins2022}.
    \item[--] So far we have discussed strong or non-existent dependencies on surface H, but in reality the mass loss could weakly depend on the surface H content. The actual dependency could perhaps be obtained by stellar atmosphere codes such as PoWR that can consistently solve the wind hydro-dynamics to determine wind parameters including mass loss. As a part of future work, a full grid of hydro-dynamically consistent PoWR could shed light on the mass-loss properties of VMS.
\end{itemize}

\section*{Acknowledgements}

We thank the anonymous referee for constructive comments that helped improve the paper. We warmly thank the MESA developers for making their stellar evolution code publicly available. JSV and ERH are supported by STFC funding under grant number ST/V000233/1.
AACS is supported by the Deutsche Forschungsgemeinschaft (DFG - German
Research Foundation) in the form of an Emmy Noether Research Group
(grant number SA4064/1-1, PI Sander).


\section*{Data Availability}

The data underlying this article will be shared on reasonable request
to the corresponding author.




\bibliographystyle{mnras}
\bibliography{References} 




\appendix


\section{Analytical expressions for $\beta (M)$, $\nabla_{\mathrm{rad}}$, $\nabla_{\mathrm{ad}}$ and $M_\mathrm{core}/M_\mathrm{star}$  }
\label{appendix:B}

Here we obtain analytical expressions for quantities required to derive the size of the convective core as a function of the initial mass. Let $M_\mathrm{init}$ be the initial mass of the star, with initial hydrogen, helium and metal mass fraction X, Y and Z respectively. In Sect. \ref{sec:methods} we described how to obtain X and Y given a Z. Let $L_\mathrm{hom} (M, X_\mathrm{s})$ be the luminosity of a fully homogeneous star with a given mass $M$ and surface hydrogen mass fraction $X_\mathrm{s}$. Fits for $L_\mathrm{hom} (M, X_\mathrm{s})$ obtained for two different mass ranges are provided below (Series 2 and 3 in Grafener 2011). For $M$ between $2\;- 100\; M_\odot$
\begin{equation}
\begin{split}
\mathrm{log}(L/L_\odot) \; = \;& (1.967 - 2.943\;X_\mathrm{s}) \\ &
+ (3.755 + 1.206\; X_\mathrm{s})\; \mathrm{log}(M/M_\odot) \\ & 
+ (-0.727  - 0.026\; X_\mathrm{s})\; \mathrm{log}(M/M_\odot)^2
\end{split}
\label{eq:L_M_X_relation1}
\end{equation}
For $M$ between $100\;- 4000\; M_\odot$, we use the following relation
\begin{equation}
\begin{split}
\mathrm{log}(L/L_\odot) \; = \;& (3.862 - 2.486\;X_\mathrm{s}) \\ &
+ (1.527 + 1.247\; X_\mathrm{s})\; \mathrm{log}(M/M_\odot) \\ & 
+ (-0.076 - 0.183\; X_\mathrm{s})\; \mathrm{log}(M/M_\odot)^2
\end{split}
\label{eq:L_M_X_relation2}
\end{equation}
The above fits evaluate the homogeneous luminosity at ZAMS given an initial mass $M_\mathrm{init}$ and an initial metallicity $Z_\mathrm{init}$.

The contribution of gas and radiation pressure to the total pressure can be quantified using $\beta$.
\begin{equation}
\begin{split}
\beta = \dfrac{P_{\mathrm{gas}}}{P_{\mathrm{tot}}}, \;\; P_\mathrm{gas}(m) = \dfrac{k\rho(m) T(m)}{\mu m_u }
\end{split}
\label{eq:gas_p}
\end{equation}
\begin{equation}
\begin{split}
1 - \beta = \dfrac{P_{\mathrm{rad}}}{P_{\mathrm{tot}}}, \;\;P_\mathrm{rad}(m) = \dfrac{a T(m)^4}{3}
\end{split}
\label{eq:rad_p}
\end{equation}
where we assume ideal gas equation of state to evaluate the gas pressure. $\rho(m)$ and $T(m)$ are the local density and temperature at location $m(r)$, $\mu$ is the mean molecular weight which can be approximated for a given Z as $\mu \approx (2X + 3/4Y + 1/2Z)^{-1}$, $k = 1.3806 \times 10^{-23} \mathrm{m}^2\; \mathrm{kg}\; \mathrm{s}^{-2}\; \mathrm{K}^{-1}$ is the Boltzmann constant, $m_u = 1.660 \times 10^{-27} \; \mathrm{kg}$ is the atomic mass unit and $a = 4\sigma/c = 7.566 \times 10^{-16} \mathrm{J}\; \mathrm{m}^{-3}\; \mathrm{K}^{-4}$ is the energy density constant. 

The virial theorem of an ideal gas connects the two energy reservoirs of the gas, the internal energy $U$ which is only a function of temperature and the gravitational potential energy $\Omega$. For a given mass distribution for a star, one can evaluate the gravitational potential energy and derive an expression for the average temperature inside the star,
\begin{equation}
\begin{split}
U  = -\dfrac{1}{2} \Omega 
\end{split}
\label{eq:Virial_thrm}
\end{equation}
\begin{equation}
\begin{split}
U  = \int u dm = \dfrac{3}{2}\int \bigg(\dfrac{P_{\mathrm{tot}}}{\rho}\bigg)dm = \dfrac{3k}{2\mu m_u} \dfrac{T_\mathrm{avg} M}{\beta_\mathrm{avg}} 
\end{split}
\label{eq:int_energy}
\end{equation}
\begin{equation}
\begin{split}
\Omega = - \alpha \dfrac{GM^2}{R}
\end{split}
\label{eq:grav_energy}
\end{equation}
Eq. \ref{eq:Virial_thrm} is the virial theorem, Eq. \ref{eq:int_energy} relates the internal energy per unit mass to the internal pressure and the appropriate average temperature and $\beta$ over the entire star, and Eq. \ref{eq:grav_energy} is the gravitational potential energy of a spherical mass distribution where $\alpha$ is determined by the actual distribution of mass inside the sphere. We take $\alpha$ value of $3/5$ corresponding to a uniform spherical mass distribution. 

Using Eq. \ref{eq:gas_p}, \ref{eq:rad_p}, \ref{eq:Virial_thrm}, \ref{eq:int_energy}, \ref{eq:grav_energy} and averaging the density over the entire star $\rho_\mathrm{avg} = 3M/4\pi R^3$, one obtains
\begin{equation}
\begin{split}
\dfrac{\beta^4}{1-\beta} \approx \dfrac{477.77}{\mu^4 M^2}
\end{split}
\label{eq:beta_mass_relation}
\end{equation}
where $M$ is in solar units. For low enough values of $\beta$, this expression reduces to $\beta \sim 1/\sqrt{M}$. A word of caution regarding the $\beta$ here, solving Eq. \ref{eq:beta_mass_relation} gives an approximate $\beta$ averaged over the entire star and is a constant for a star of given mass. While $\beta$ in Eq. \ref{eq:gas_p} is a function of $m(r)$ and in general varies inside the star. 

\begin{figure}
    \includegraphics[width = \columnwidth]{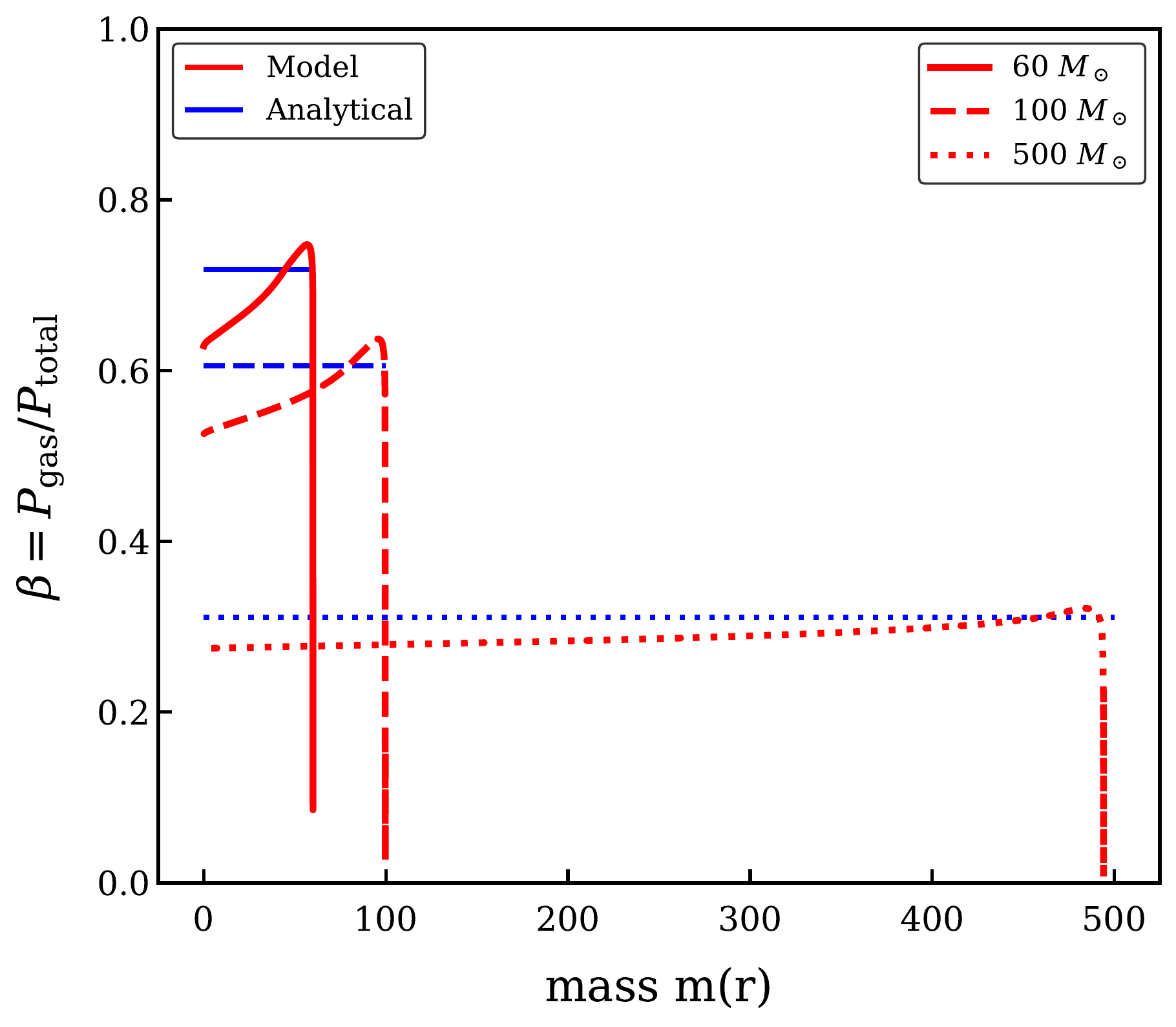}
    \caption{Comparison of the value of $\beta = P_\mathrm{gas}/P_\mathrm{tot}$ inside the star obtained from detailed stellar evolution models (red) with that obtained analytically (blue). }
    \label{fig:beta_model_theory}
\end{figure}

For a given star with initial mass $M_\mathrm{init}$ and initial metal mass fraction $Z_\mathrm{init}$, one can solve Eq. \ref{eq:beta_mass_relation} for $\beta$. In Fig. \ref{fig:beta_model_theory}, we plot the variation of $\beta$ at ZAMS for three representative stellar models with $M_\mathrm{init} = 60, 100$ and $500 \; M_\odot$ at Galactic metallicity and compare it with $\beta$ obtained analytically. The quantitative accuracy of $\beta_\mathrm{avg}$ depends on various assumptions and simplifications made in its derivation, especially the `average' that was performed in Eq. \ref{eq:int_energy}. This `averaging' over the entire star causes a systematic offset of $\approx 0.1$ in the value of $\beta$ near the core of the star. Nevertheless there is a good qualitative agreement between the models and the $\beta_\mathrm{avg}$ predicted from the analytical expression, both decreasing with increasing initial mass. 

\begin{figure}
    \includegraphics[width = \columnwidth]{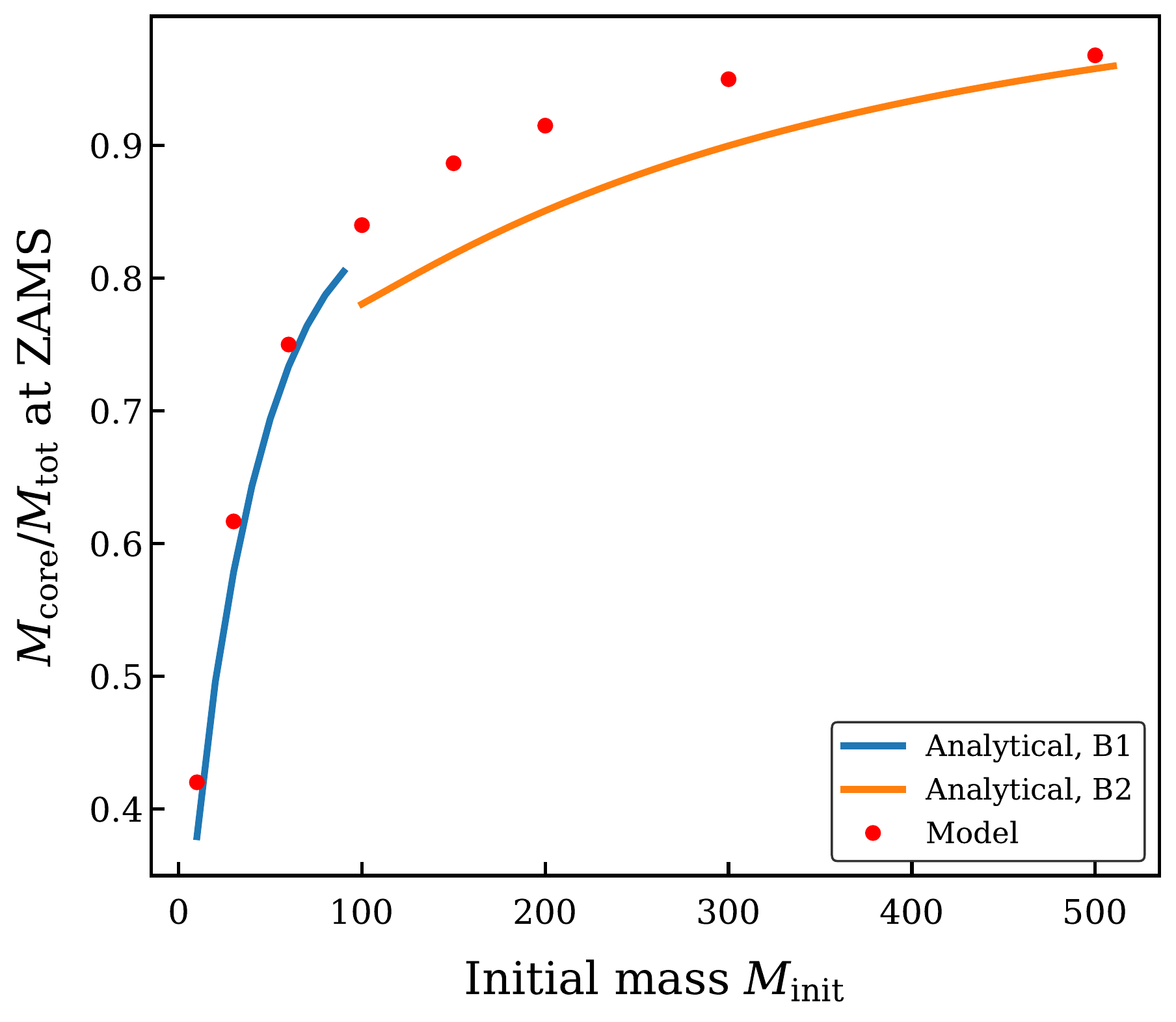}
    \caption{Comparison of convective core-to-total mass ratio at ZAMS as a function of initial mass obtained from detailed stellar models and from analytical expressions.}
    \label{fig:zams_core_size}
\end{figure}

The convective core boundary is defined at the location $m(r)$ where $\nabla_{\mathrm{rad}} = \nabla_{\mathrm{ad}}$. The adiabatic temperature gradient describes the temperature changes in a mass element undergoing an adiabatic transformation (expansion or contraction). Using the first law of thermodynamics and basic thermodynamic relations for specific heats, $\nabla_{\mathrm{ad}}$ of a layer of star supported by both gas and radiation pressure is given as
\begin{equation}
\begin{split}
\nabla_{\mathrm{ad}} = \dfrac{\beta^2 + (1-\beta)(4+\beta)}{2.5\beta^2 + 4(1-\beta)(4+\beta)}
\end{split}
\label{eq:ad_beta_mass_relation}
\end{equation}
For low values of $\beta$ (higher fraction of radiation pressure), this reduces to $\nabla_{\mathrm{ad}} \approx 0.25$.

The quantity $\nabla_{\mathrm{rad}}$ describes the temperature gradient required inside the star to transport the entire stellar luminosity by radiation diffusion, and is given by
\begin{equation}
\begin{split}
\nabla_{\mathrm{rad}} = \dfrac{3}{16\pi acG} \dfrac{\chi l(m)P(m)}{mT(m)^4} = \dfrac{1}{16\pi cG}\dfrac{\chi l(m)}{m (1-\beta)}
\end{split}
\label{eq:rad_beta_mass_relation}
\end{equation}
The location of switch from convective core to the radiative envelope is well outside the burning region and one can take $l(m) \approx L$. The opacity in these layers can also be approximated by the electron scattering opacity, $\chi = \chi_e = 0.02 (1+X)\; \mathrm{m}^2\; \mathrm{kg}^{-1}$. Given the solution for $\beta$ obtained earlier and $L$ for fully homogeneous stars in Eq. \ref{eq:L_M_X_relation1} and \ref{eq:L_M_X_relation2}, one can solve for $m(r)$ where $\nabla_{\mathrm{rad}} = \nabla_{\mathrm{ad}}$. 
\begin{equation}
\begin{split}
M_\mathrm{core} = 6.47 \times 10^{-6} \Bigg(\dfrac{L/L_\odot}{(1-\beta)\nabla_\mathrm{ad}}\Bigg)
\end{split}
\label{eq:m_core_size}
\end{equation}
The quantitative accuracy of $M_\mathrm{core}$ relies on both the luminosity and the value of $\beta$ evaluated. Fig. \ref{fig:zams_core_size} compares the value of core to total mass at ZAMS as a function of the initial mass from MESA models to that obtained analytically. Remember the core size in massive star models, atleast below the transition point is very sensitive to the core overshooting and rotation physics included. Nevertheless we find a good agreement between the core size predicted from models and analytically throughout the entire mass range for the luminosity considered here. Above $M_\mathrm{init} \sim 200 \; M_\odot$, the core encompasses almost $90\%$ of the entire star, resulting in chemically homogeneous abundance profile throughout the star.

\bsp	
\label{lastpage}
\end{document}